\documentclass[12pt]{article}
\usepackage{amsmath,amssymb,amsfonts,epsfig}
\usepackage[nosort]{cite}

\makeatletter\@addtoreset{equation}{section}\makeatother

\newcommand{\Tr}{\text{Tr}\ }

\newcommand{\nn}{\nonumber }

\newcommand{\bR}{\mathbb{R} }

\newcommand{\bC}{\mathbb{C} }

\allowdisplaybreaks

\input epsf

\begin{document}

\thispagestyle{empty}
\vspace*{-2em}
\begin{flushright}
UT-KOMABA/08-22\\ 
December 2008
\end{flushright}
\vspace{0.3cm}
\begin{center}
\Large {\bf
Boundary condition for D-brane from Wilson loop, \\
and gravitational interpretation of eigenvalue \\
in matrix model in AdS/CFT correspondence}

\vspace{0.7cm}

\normalsize
 \vspace{0.4cm}

Shoichi {\sc Kawamoto}$^{a,}$\footnote{e-mail address:\ \ 
{\tt kawamoto@ntnu.edu.tw}}
,
Tsunehide {\sc Kuroki}$^{b,}$\footnote{e-mail address:\ \ 
{\tt tkuroki@rikkyo.ac.jp}}
and
Akitsugu {\sc Miwa}$^{c,d,}$\footnote{e-mail address:\ \ 
{\tt akitsugu@hep1.c.u-tokyo.ac.jp}}

\vspace{0.7cm}

$^a$ 
{\it Department of Physics,
National Taiwan Normal University,\\
88, Sec.4, Ting-Chou Road,
Taipei 11650, Taiwan.}\\

\vspace{0.4cm}

$^b$
{\it 
Department of Physics, Rikkyo University, \\
Tokyo 171-8501, Japan.}\\

\vspace{0.4cm}

$^c$
{\it Institute of Physics, University of Tokyo,\\
Komaba, Meguro-ku, Tokyo 153-8902, Japan.}\\

\vspace{0.4cm}
$^d$
{\it Harish-Chandra Research Institute,\\
Chhatnag Road, Jhunsi, Allahabad 211019, India.}\\

\vspace{1cm}
{\bf Abstract}
\end{center}
We study the supersymmetric Wilson loops in the four-dimensional ${\cal
 N}=4$ super Yang-Mills theory in the context of AdS/CFT
correspondence.
In the gauge theory side, it is known that the expectation value of
the Wilson loops of circular shape with winding number $k$, $W_k(C)$,
is calculable by using a Gaussian matrix model.
In the gravity side,
the expectation value of the loop is conjectured to be
given by the classical value of the action
$S_{\rm D3}$ for a probe D3-brane with $k$ electric fluxes
as $\langle W_k(C) \rangle = {\rm e}^{-S_{\rm D3}}$\,.
Given such correspondence, we pursue the interpretation
of the matrix model eigenvalue density, or more precisely the resolvent,
from the viewpoint of the probe D3-brane.
We see that the position of an eigenvalue appears
as an integrated flux on the D3-brane.
In the course of our analysis, we also clarify
the boundary condition on the D3-brane in terms of
the Wilson loop.

\vspace{2em}
{\noindent
PACS codes:
11.15.Pg, \,
 11.25.Tq, \,
11.25.Uv. \,
\\[0.5em]
Keywords: AdS/CFT correspondence, Wilson loops, matrix models.
}
\newpage
\setcounter{page}{1}

\tableofcontents

\setcounter{footnote}{0}

\section{Introduction}
\label{sec:intro}

Duality between string theory and gauge theory has been 
an important concept for theoretical particle physics. 
In particular Wilson loop would play a unique role there, 
since originally it has been introduced in the context 
of stringy behavior of the strong coupling gauge theory, 
i.e., the area law behavior in the confining gauge theory.

In the recent development of the duality, 
the ``area law'' of the Wilson loop 
has been rediscovered \cite{holographicWL} 
in the context of the AdS/CFT correspondence. 
Hence in AdS/CFT, Wilson loop is useful for concrete realization 
of the duality. There, we usually begin with the $N$ D3-branes 
located on top of each other and put another D3-brane
in parallel with them at a large distance. 
Then we evaluate the amplitude for the propagation of a string 
stretched between the $N$ D3-branes and the single separated brane
from two different points of view, from gauge theory 
and from gravity picture.

From the viewpoint of the gauge theory, the stretched string 
corresponds to a bi-fundamental matter of the SU($N$)$\times$U($1$) 
gauge theory. 
By evaluating the amplitude for the string propagation along a loop $C$ 
on the isolated brane, we can introduce a Wilson loop for $C$ 
in the bi-fundamental representation
of the SU($N$)$\times$U($1$) gauge group.

In order to discuss the gravity side, we replace the $N$ D3-branes 
with the near horizon geometry of the extremal black 3-brane solution. 
Then the isolated brane is recognized as
a single probe D3-brane in the near horizon region and we 
evaluate the propagation amplitude for the string attached to the 
loop $C$ on the probe D3-brane. 

The AdS/CFT correspondence claims that 
these two different points of view give the 
same result. Therefore we reach the following conjecture:
\begin{align}
\bigg \langle
\frac{1}{N}
 {\rm tr\, P\!}
\exp
\Big(
\int ds 
( i A_\mu \dot x^\mu + |\dot x| \theta_i \Phi_i )
\Big)
\bigg \rangle_\text{CFT}
=
\int_C dX {\rm e}^{-S_\text{string}}\,. \label{RY-M}
\end{align}
Here on the left hand side, not only the gauge field 
but the scalar fields are included since a string 
attached to D-branes is coupled to all of them. 
Also, the U($1$) part of the Wilson loop has been omitted 
since we discuss the limit in which the U($1$) brane is 
separated by a large distance from the remaining $N$ D3-branes
and thus its dynamics decouples from the SU($N$) part we are interested in.
Actually in the gravity side we consider the string world sheet 
which is attached to the loop $C$ on the AdS boundary.
Hence the right hand side only takes account of the SU($N$) 
contribution in terms of the gauge theory side.
Also we need to take care of the definition of the 
functional $S_\text{string}$ on the right hand side.
In the paper \cite{Drukker:1999zq}
it has been proposed that in addition to the
usual Nambu-Goto type action, we need to add appropriate 
boundary terms which eliminate 
divergence due to the infinite
scale factor from the vicinity of the AdS boundary.
Introducing these boundary terms was recognized as performing the
Legendre transformation to change the boundary condition of the
world sheet.

The conjecture \eqref{RY-M} has been checked mainly 
in the case of the straight Wilson line or 
the circular Wilson loop which preserves some global
supersymmetry.\footnote{
For a recent quantitative test of a similar conjecture to \eqref{RY-M}
in the case with less symmetry,
see the paper \cite{Hanada:2008gy}. 
There a similar relation in the case of 
finite temperature D0-brane system
is tested by using the Monte Carlo simulation.}
In the study of the circular Wilson loop, 
a Gaussian matrix model plays an important role.
It was proposed as a technical tool
for summing up all the planar ladder 
diagrams \cite{Erickson:2000af} and 
a further argument 
for the matrix model including the non-planar diagrams 
was given in \cite{Drukker:2000rr}.\footnote{%
See also \cite{Chu:2007pb} for a discussion on the relation
between the Wilson loop and the matrix model
in the beta-deformed super Yang-Mills theory.}
In the more recent paper \cite{Pestun:2007rz}, an argument 
based on the topological twist and localization technique is discussed.
A relation to the Gaussian matrix model has
also been discussed by using the mirror symmetry and the geometric
transition\cite{Bonelli:2008rv}\,.

In the study of the validity of the conjecture \eqref{RY-M},
an important work has been done in the paper \cite{Drukker:2005kx}.
There the authors considered a circular Wilson loop 
with winding number $k$. 
This winding number corresponds to the string charge 
because $k$ stretched strings between $N$ D3-branes 
and the single D-brane mentioned above yield the Wilson loop 
with winding number $k$. 
In the gauge theory side the expectation value of the operator
can be evaluated by the matrix model of the same Gaussian action
as in the single winding case.
On the other hand, in the gravity side, 
they considered a spike D3-brane solution carrying a non-trivial electric flux 
by $k$ units of the electric charge, 
because the electric charge on the D-brane corresponds to the string charge. 
In the gravity side the computation is essentially the 
same as the right hand side of \eqref{RY-M} with 
the string action being replaced by
the D3-brane action.\footnote{
It has been proposed that the D3-brane actually corresponds 
to the Wilson loop in the symmetric representation 
\cite{Gomis:2006sb},
and has been argued that in the strong coupling limit of the gauge
theory, a multiply wound Wilson loop and a symmetric one give 
an identical expectation value.
See the paper \cite{Okuyama:2006jc}, for these issues.
We however consider that there is a subtlety here. See the concluding section.} 
In \cite{Drukker:2005kx}, it was found 
that the relevant D3-brane solution has the geometry of 
AdS$_2 \times$S$^2$ where the radius of the S$^2$ is given 
by the parameter $\kappa \equiv k \sqrt{\lambda} /(4N)$.
This suggests that the large-$N$ limit with fixed $\kappa$
in the gauge theory side is a quite interesting limit.
Actually it was found that in this limit, 
the expectation value of the Wilson loop 
computed by using the Gaussian matrix model
agrees with the prediction of the gravity side.
A very important point in this work was that 
the large-$N$ limit is not the planar limit but 
the result contains a class of the non-planar contributions 
since the parameter $\kappa$ depends on $N$ inversely.

One of the interesting developments which followed 
the paper \cite{Drukker:2005kx} is the study
of the geometric aspects of the eigenvalue in the 
Gaussian matrix model.
Stimulated by the case of a local operator \cite{Lin:2004nb},
the correspondence between the eigenvalue distribution
and a certain aspects of the geometry has been 
discussed in \cite{Yamaguchi:2006te,Lunin:2006xr}. 
In general, gravitational interpretation of eigenvalues in matrix models 
would be an interesting and essential problem for gauge/string duality, 
or even for nonperturbative formulation of string theory. 
For example, in the IIB matrix model\cite{Ishibashi:1996xs} 
the eigenvalues of matrices 
are interpreted as the space-time points 
and their dynamics is discussed as emergent geometry
\cite{Nishimura:2001sx}. 
Therefore, even if we concentrate on a particular eigenvalue 
of a specific matrix model, it would be intriguing to find 
its gravitational interpretation clearly in the context of the AdS/CFT correspondence.

Information on the eigenvalue of a matrix model is 
well packed in the expectation value of an operator called the
resolvent.
In this paper we thus consider the resolvent 
of the matrix model from the viewpoint of the probe D3-brane.
Our main goal is to discuss a gravitational 
interpretation of the eigenvalue distribution, or more precisely
the resolvent, in the Gaussian matrix model in the presence of the 
Wilson loop with a large winding number $k$.
In particular we are interested in results which follow directly 
from the basic correspondence \eqref{RY-M} using the D3-brane action.
This is in contrast to the ``bubbling geometry''
approach taken in \cite{Lin:2004nb,Yamaguchi:2006te,Lunin:2006xr}. 
On the way of the analysis, we also identify what kind of boundary conditions
should be imposed on the D3-brane configuration so that it will correspond to
the Wilson loop with winding number $k$.

The rest of the paper is organized as follows.
In section \ref{sec:WLGT}, we review the matrix model 
computation and discuss the resolvent in the $1/N$-expansion 
with fixed $\kappa$. 
From this result, we will see what kind of quantity 
in the gravity side should 
reproduce the matrix model resolvent.
In section \ref{sec:D3-Wilsonloop}, 
we begin with a brief review of the 
setup and the results of \cite{Drukker:2005kx}.
Next we perform the computation in the gravity side 
which yields the same result as in the matrix model,
and then we find a gravitational interpretation of an eigenvalue.
In the course of the analysis 
we will give some interesting aspects 
of the work \cite{Drukker:2005kx}, especially identification of 
appropriate boundary conditions for D3-brane configurations. 
We then conclude the paper in section \ref{sec:conclusion}
with some discussions.
A couple of appendices are devoted to filling in some technical details
in the main part of the paper.

\section{Multiply-wound Wilson loops in gauge theory}
\label{sec:WLGT}

In this section we discuss the gauge theory side and 
its relationship to the matrix model. 
In subsection \ref{sec:matrixmodel}, we start with a brief review
of the  expectation value of the Wilson loop  
in the large-$N$ matrix model.
We also mention the eigenvalue distribution derived in 
\cite{Yamaguchi:2006te} by solving the saddle point equation 
of the matrix model. 
For our present purpose, however, it will turn out to be more 
useful to discuss the resolvent through the Laplace transformation.
In subsection \ref{sec:Laplace} we discuss the Laplace
transformation that motivates
 us to proceed to the gravitational interpretation.

\subsection{Matrix model calculation}
\label{sec:matrixmodel}

We consider the Wilson loop  
in the four-dimensional ${\cal N}=4$ super Yang-Mills theory,
which is defined by
\begin{equation}
 W_k(C) = \frac{1}{N} {\rm tr\,P\!}\exp
\bigg( 
\int ds 
\Big(i A_\mu(x(s)) \dot x^\mu (s) + \Phi_i(x(s)) |\dot x(s)| \theta^i(s) \Big)
\bigg)\,.
\label{WilsonLoop}
\end{equation}
Here $A_\mu(x)$ ($\mu=1\ldots4$) and 
$\Phi_i(x)$ ($i=1\ldots6$) are the gauge field 
and the scalar fields, respectively.
The coordinates $x^\mu$ span the Euclidean 
four-dimensional space and $\theta^i$ is the 
coordinate on the unit S$^5$\,.
We consider the operator in which 
the path $C:\{x^\mu (s) \,| \, 0\leq s < 2 k \pi\}$
is a multiply winding circle and $\theta^i$ is constant.
The subscript $k$ on the left hand side 
indicates that the loop $C$ goes around 
the circle $\{{x^\mu(s) \, | \, 0 \leq s < 2 \pi}\}$ 
$k$ times.

The expectation value of the operator 
\eqref{WilsonLoop} is calculable by 
means of the Gaussian matrix model
\cite{Erickson:2000af,Drukker:2000rr,Pestun:2007rz}:
\begin{align}
\big \langle W_k(C) \big \rangle_{\mathcal{N}=4 \, \text{SYM}} 
&=\bigg \langle \frac{1}{N} {\rm tr}\,{\rm e}^{kM} \bigg
\rangle_\text{matrix model}
\nn\\
&=
{1 \over Z}
\int dM {1 \over N} {\rm tr}\,{\rm e}^{kM} 
{\rm e}^{ - {2 N \over \lambda} {\rm tr}\,M^2}
=
{ 1 \over N } 
{\rm e}^{{k'}^2 \over 2} L_{N-1}^{(1)} ( - {k'}^2  )
\equiv f(k',N)
\label{W=Laguerre} 
\,, \\
Z 
&\equiv \int d M {\rm e}^{ - {2 N \over \lambda} {\rm tr}\,M^2}\,.
\label{Z}
\end{align}
Here $k' \equiv \sqrt{\lambda /(4 N)}k$ and 
$
L_n^{(\alpha)}(\zeta)
\equiv 
({\rm e}^{\zeta} \zeta^{-\alpha}/n!)
d^n/d \zeta^n
(e^{-\zeta} \zeta^{n+\alpha})
$
is the Laguerre polynomial.
We introduced the function $f(k',N)$ for later convenience.
Hereafter, the expectation value of the Wilson loop 
(\ref{WilsonLoop}) is always understood as
this matrix model expectation value.

In \cite{Drukker:2005kx}, the large-$N$ limit of $f(k',N)$ 
with fixed $\kappa \equiv k \sqrt{\lambda}/(4N)$ 
was derived by using a differential equation.
Here we follow their steps and introduce 
${\cal F}(\kappa,N)\equiv -N^{-1} 
\log  f(k',N)
$\,.
Then ${\cal F}$ satisfies the following 
differential equation:
\begin{equation}
( \partial_\kappa {\cal F} )^2
-{ 1 \over \kappa N}
( \kappa \partial_\kappa^2 {\cal F} 
+ 3 \partial_\kappa {\cal F})
-
16 ( 1+\kappa^2)
=0\,.
\label{Deq}
\end{equation}
In the large-$N$ limit with fixed $\kappa$, 
the differential equation \eqref{Deq} can 
be solved perturbatively and the solution is given by\footnote{
The constant term
can be fixed, for example,
by comparing the small $\kappa$ limit of \eqref{cal F}
with the modified Bessel function.
For the relation between the function $f(k',N)$ with small $\kappa$
and the modified Bessel function, 
see the discussion around \eqref{modified-Bessel}.}
\begin{align}
& {\cal F_{\pm}}(\kappa,N)  \notag \\
& =
\pm
2 
\Big( \kappa \sqrt{ 1 + \kappa^2 } 
+ 
{\rm arcsinh} \kappa \Big) 
+
{1 \over 2 N}
\bigg(
\log \kappa^3 \sqrt{1+\kappa^2} 
+
\log(32 \pi N^3)
\bigg)
+
{\cal O}( N^{-2})\,.
\label{cal F}
\end{align}
We take $\mathcal{F}_-$ since it dominates in the large-$N$ limit.

Now let us see that 
the leading term in (\ref{cal F}) originates 
from an eigenvalue apart from 
the cut of the standard semi-circle distribution 
of the Gaussian matrix model \cite{Yamaguchi:2007ps}. 
In terms of eigenvalues, (\ref{W=Laguerre}) can be written as 
\begin{align}
&\big \langle W_k(C) \big \rangle 
={1 \over Z}\int \prod_i dm_i\exp(-V_\text{eff}), \notag \\
&V_\text{eff}\equiv N\sum_{i=1}^{N-1} V(m_i)-\sum_{i>j}^{N-1}\log(m_i-m_j)^2
       +NV(m_N)-\sum_{j=1}^{N-1}\log(m_N-m_j)^2-km_N, \notag\\
&V(m)=\frac{2}{\lambda}m^2,
\end{align} 
which tells us that the Wilson loop gives rise to force 
only on a single eigenvalue (say $m_N$).  
Therefore let us introduce the eigenvalue distribution as
\begin{align}
\rho(m)&=\rho^{(0)}(m)+\frac{1}{N}\rho^{(1)}(m), \notag \\
\rho^{(0)}(m)&={1 \over N}\sum_{i=1}^{N-1}\delta(m-m_i), \qquad
\rho^{(1)}(m)=\delta(m-m_N). \label{rho0-rho1}
\end{align} 
Then $V_\text{eff}$ can be rewritten as 
\begin{align}
V_\text{eff} 
& =N^2\int dm\rho^{(0)}(m)V(m)-\frac{N^2}{2}\int dmdm'\rho^{(0)}(m)\rho^{(0)}(m')\log(m-m')^2 \notag\\
                 & \hspace{-0.3cm}+ N\int dm\rho^{(1)}(m)V(m)-N\int dmdm'\rho^{(0)}(m)\rho^{(1)}(m')\log(m-m')^2-k\int dm \rho^{(1)}(m)m.
\label{Veff}
\end{align}
By considering the variation of $\rho^{(0)}(m)$,
the saddle point equation reads 
\begin{align}
V'(m)- 2\int \hspace{-0.4cm}-  
dm'\frac{\rho^{(0)}(m')}{m-m'}
- {2 \over N} \int \hspace{-0.4cm}- 
dm'\frac{\rho^{(1)}(m')}{m-m'}=0.
\label{rho-0-saddle}
\end{align}
In order for the subleading distribution $\rho^{(1)}(m)$ 
to make sense, we need to discuss
the distribution $\rho^{(0)}(m)$ to the same order.
Hence we further expand it with respect to $1/N$ as
\begin{equation}
\rho^{(0)}(m) = \rho^{(0,0)}(m) + {1 \over N} \rho^{(0,1)}(m)
+ {\cal O}(N^{-2}). 
\label{decomposition}
\end{equation}
Here we note that the distribution functions 
$\rho^{(0,0)}(m)$ and $\rho^{(0,1)}(m)$ satisfy 
the following conditions:
\begin{equation}
\int dm \rho^{(0,0)}(m) = 1, \quad 
\int dm \rho^{(0,1)}(m) = -1,
\label{int dm rho}
\end{equation}
which can be seen from \eqref{rho0-rho1}
and the expansion \eqref{decomposition}.
Hence, formally, $\rho^{(0,0)}(m)$ corresponds 
to the distribution of $N$ eigenvalues while
$(1/N) \rho^{(0,1)}(m)$ subtracts a single eigenvalue. 
By taking terms of order $N^0$ in \eqref{rho-0-saddle}, 
we find that $\rho^{(0,0)}(m)$ satisfies the 
saddle point equation for the semi-circle distribution
$2\pi \rho^{(0,0)}(m)=\sqrt{\frac{16}{\lambda}-\frac{16}{\lambda^2}m^2}$
with support $(-\sqrt{\lambda},\sqrt{\lambda})$.

Next, before discussing $\rho^{(0,1)}(m)$,
we shall study the saddle point equation which will be 
derived by considering the variation of $\rho^{(1)}(m)$.
In order to solve this,
it is reasonable to assume that $m_N$ 
is isolated from the other $N-1$ eigenvalues
due to the extra force coming from the Wilson loop and to
make an ansatz $\rho^{(1)}(m)=\delta(m-m_*)$
with $m_\ast$ outside the range $(-\sqrt{\lambda},\sqrt{\lambda})$.
Then the saddle point equation for positive $m_*$ becomes  
\begin{align}
V'(m_*)-\left(V'(m_*)-\sqrt{V'(m_*)^2-\frac{16}{\lambda}}\right)
-\frac{k}{N}=0,
\label{spe}
\end{align} 
from which we find 
\begin{align}
m_*=\sqrt{\lambda}\sqrt{1+\kappa^2}, 
\label{isolated}
\end{align}
which is indeed outside the cut.\footnote
{This is the unique solution for $k>0$. 
For negative $k$, we would have
the same solution with the opposite sign.
In this paper we assume that $k>0$ without loss of generality.}

Finally, we discuss $\rho^{(0,1)}(m)$.
Let us first
study the behavior of $\rho^{(0,1)}(m)$
in the limit $\kappa \to \infty$ by a physical argument
without using the saddle point equation.
In this limit, the isolated eigenvalue goes to infinity
and its effect on the remaining $N-1$ eigenvalues 
should vanish. 
Then the resulting distribution function $\rho^{(0)}(m)$
becomes the semi-circle distribution for $N-1$ eigenvalues 
which is given by
\begin{align}
\lim_{\kappa \to \infty} \rho^{(0)}(m) 
& = 
{N-1 \over N} \times 
{2 \over \pi} {1 \over \lambda_{N-1}}
\sqrt{\lambda_{N-1} - m^2} + {\cal O}(N^{-2}) 
\label{kappa->inf1}\\
&= 
{2 \over \pi} {1 \over \lambda}
\sqrt{\lambda - m^2}
-
{1 \over N}
{1 \over \pi}
{1 \over \sqrt{\lambda - m^2}}
+ {\cal O}(N^{-2}).
\label{kappa->inf}
\end{align}
Here we have introduced $\lambda_{N-1} = (N-1)/N \times \lambda$
and the overall factor $(N-1)/N$ on the right hand side 
of \eqref{kappa->inf1} reflects the fact that $\rho^{(0)}(m)$
is defined with the overall factor $1/N$ instead of $1/(N-1)$
(see \eqref{rho0-rho1}).
So, from this simple argument we find that, 
in the limit $\kappa \to \infty$, $(1/N)\rho^{(0,1)}(m) $
is given by the second term of \eqref{kappa->inf} 
(including the minus sign).

For a generic value of $\kappa$, we need to solve the
saddle point equation 
for $\rho^{(0,1)}(m)$, i.e., the terms of ${\cal O}(N^{-1})$ 
in \eqref{rho-0-saddle}, which is now given by
\begin{equation}
- \int \hspace{-0.4cm}-  
dm'\frac{\rho^{(0,1)}(m')}{m-m'} 
-
{ 1 \over m - m_\ast}
=
0. \label{rho-01-saddle}
\end{equation}
Physically, this equation shows that 
the subleading distribution $\rho^{(0,1)}(m)$ 
is determined by the following manner:
the repulsive force from the 
isolated eigenvalue at $m=m_\ast$ changes
the subleading distribution 
$\rho^{(0,1)}(m)$ of the $N-1$ eigenvalues 
from its form in the limit $\kappa \to \infty$
discussed above.
This change gives rise to additional 
force among the $N-1$ eigenvalues themselves.
Then the distribution $\rho^{(0,1)}(m)$ is
determined by the force balance condition among
these effects.
Therefore, $\rho^{(0,1)}(m)$ describes both effects 
of subtracting a single eigenvalue and distortion 
on the distribution of the remaining eigenvalues 
due to the isolated eigenvalue. 
It is evident that they are of ${\cal O}$(1/N). 
Since $\rho^{(0,1)}(m)$ subtracts an eigenvalue, 
it is natural to assume that it has its support only inside the 
range $(-\sqrt{\lambda},\sqrt{\lambda})$ or at $m=m_\ast$. 
This is because the total eigenvalue density $\rho(m)$
must be positive everywhere. 
In fact, as we have shown above, 
the support of $\rho^{(0,1)}(m)$ in the limit 
$\kappa \to \infty$ is given by $(-\sqrt{\lambda},\sqrt{\lambda})$. 
Furthermore, since the distortion due to the isolated eigenvalue 
should reach the whole distribution of the remaining eigenvalues, 
it is natural again that the support of $\rho^{(0,1)}(m)$ is 
$(-\sqrt{\lambda},\sqrt{\lambda})$ for a generic value of $\kappa$. 
In the rest of this subsection we assume this property.\footnote{
In the next subsection and 
appendix \ref{sec:eval-secondterm}, 
we give a rigorous discussion of this model 
by using the orthogonal polynomial method 
without making any assumption. 
In particular, in appendix \ref{sec:eval-secondterm}
we derive the explicit form of the resolvent independently,
which also provides a proof of this property.}

Let us solve the saddle point equation \eqref{rho-01-saddle}
based on this physical argument.
For this purpose, we introduce the resolvent $R^{(0,1)}(m)$ as
\begin{equation}
R^{(0,1)}(m) 
= 
{1 \over N} \int dm' {{\rho^{(0,1)}(m')} \over m - m'}\,.
\label{R(0,1)def}
\end{equation}
Then $R^{(0,1)}(m)$ is determined by the following conditions:
\begin{itemize}
\item The real part of $R^{(0,1)}(m)$ at the 
support of $\rho^{(0,1)}(m)$ is given by \eqref{rho-01-saddle}.
\item For the generic value of $\kappa$, the resolvent has one cut  
$(-\sqrt{\lambda}, \sqrt{\lambda})$ 
while it has no pole.
This is due to the assumption for the support of $\rho^{(0,1)}(m)$
mentioned above.
\item In the large-$m$ limit, the condition
$R^{(0,1)}(m) \to - {1 \over N}{1 \over m}$ should be satisfied.
This is due to the second equation of \eqref{int dm rho}.
\end{itemize}
By taking account of these conditions, 
we can fix the resolvent and the eigenvalue distribution completely as
\begin{align}
R^{(0,1)}(m) 
& = 
{1 \over N}
\bigg(
-{1 \over m - m_\ast} 
+ {\sqrt{\lambda} \kappa \over (m-m_\ast) \sqrt{m^2 - \lambda }
}
\bigg), \label{R(0,1)}\\
\rho^{(0,1)}(m)
& = 
- {N \over 2 \pi i}
\Big( R^{(0,1)}(m + i \epsilon) - R^{(0,1)}(m - i \epsilon) \Big) 
\\[2mm]
& =
\begin{cases}
& 
{\displaystyle {1 \over  \pi }
{\sqrt{\lambda} \kappa \over (m-m_\ast) \sqrt{ \lambda - m^2 }}
\quad 
(- \sqrt{\lambda} \leq m \leq \sqrt{\lambda}) \qquad  (\kappa \neq 0), }\\[5mm]
& 
- \delta \big( m - \sqrt{\lambda} \big) \hspace{5.55cm} (\kappa = 0).
\end{cases} \label{rho(0,1)}
\end{align}
Here in the case of $\kappa=0$, we have the delta function distribution
which exactly cancels the another subleading distribution
$\rho^{(1)}(m)$. 
This is indeed the expected behavior since for $\kappa=0$, 
the distribution should reduce to the leading one 
$\rho^{(0,0)}(m)$. 
It is also easy to check that in the limit $\kappa \to \infty$,
the above distribution shows the behavior of ($N$ times) 
the second term in \eqref{kappa->inf}.

Having studied the eigenvalue distribution,
we now consider the saddle point value of 
$V_{\rm eff}$. 
The leading term of order $N^2$ is just 
canceled by the same term from the denominator $Z$.
The subleading term of order $N$
which comes from the first line of \eqref{Veff}
is given by 
\begin{equation}
V_{\rm eff}^{(1),\textrm{1st line}}
=
N \int dm \rho^{(0,1)}(m)
\bigg(
V(m) - 
\int dm' 
\rho^{(0,0)}(m')
\log(m-m')^2
\bigg).
\label{addVeff1}
\end{equation}
By using the saddle point equation,
it is easy to show that the combination in 
the round bracket, which is obviously 
$k$-independent, is also $m$-independent.
From this, we find that 
\eqref{addVeff1} is just a $k$-independent constant, 
since $m$-integral of $\rho^{(0,1)}(m)$ is $k$-independent.

Next, the order $N$ term coming from the second line 
of \eqref{Veff} is given by
\begin{align}
V_\text{eff}^{(1),\textrm{2nd line}}
=N\left(-2\kappa\sqrt{1+\kappa^2}-2\text{arcsinh}\kappa 
+ {\rm constant }\right),
\label{Veff1}
\end{align}
where the last term is a $\kappa$-independent constant.
We can check that the constant term in \eqref{Veff1}
is canceled by \eqref{addVeff1}
due to the relation $\rho^{(0,1)}(m) + \rho^{(1)}(m) =0 $ 
for $\kappa=0$, which has been mentioned after \eqref{rho(0,1)}.
Therefore, we have derived the first term 
of \eqref{cal F} with confirming again 
that we should take the minus sign there.
An interesting point in this analysis is that, 
since we consider the large $k \sim N$ limit, 
the effects of the Wilson loop on 
the saddle point are not negligible; 
the standard semi-circle eigenvalue distribution 
is corrected by the $1/N$ terms
and these corrections affect the leading behavior 
of the Wilson loop expectation value.
In particular, the important contribution 
of \eqref{Veff1} originates from the 
isolated eigenvalue.\footnote
{Such isolated eigenvalues play a crucial role 
in nonperturbative effects in noncritical string theories\cite{David:1990sk,Hanada:2004im}. 
It would be interesting if the Wilson loop with winding number 
of ${\cal O}(N)$ is shown to 
play a similar role in critical string theories.}

\subsection{
 Resolvent 
through the Laplace transformation}
\label{sec:Laplace}

The eigenvalue distribution discussed in the previous subsection 
is a kind of master field for the correlators in the presence of the 
Wilson loop. If AdS/CFT can be regarded as realizing
the idea of the master field, it is natural to expect that 
there should be a counterpart of the eigenvalue distribution 
in the gravity side. Here we note that in general multi-matrix models 
like the four-dimensional ${\cal N}=4$ super Yang-Mills theory would not 
have a master field, but in a certain BPS sector like circular Wilson loops 
a kind of master field may exist due to higher supersymmetries 
by which the system can be effectively described by a one-matrix model.    
As we have already used in the previous section, 
the eigenvalue distribution is given by
the imaginary part of the resolvent, $R(z)$\,, 
which is more tractable as a complex function, 
and the position of the eigenvalue is the pole of $R(z)$.
We thus study how the isolated pole arises 
when we discuss the resolvent based on 
its standard definition:
\begin{align}
R(z) 
&=
{1 \over Z_V}\int dM {1 \over N} {\rm tr}{1 \over z - M} {\rm e}^{-V(M)}\\
&=
\int_0^\infty dp \, {\rm e}^{-pz}
{
{1 \over Z_V}
\int dM 
{1 \over N}
{\rm tr}\, {\rm e}^{pM}
{\rm e}^{-V(M)} 
} \,, \label{R(z)} \\
Z_V &= \int dM {\rm e}^{-V(M)}.
\end{align}
Here $V(M)$ is a potential whose saddle point
we are interested in.
Such a question may be interesting even 
purely from the matrix model viewpoint.
What is more, it will guide us to a gravitational interpretation,
as we shall see.

In the integral representation \eqref{R(z)},
we should pay attention to the range of the variable $p$\,.
Since we are interested in the saddle point which is 
defined solely by the potential $V(M)$\,, 
we need to drop effects of the inserted operator 
${\rm tr}\,{\rm e}^{pM}$ on the saddle point.
This means that although the upper limit of the 
$p$-integral is set to be infinity, we still assume 
the relation $p \ll N$\,. 
In order to make this point clear,
let us recall the case of the Gaussian potential,
$V(M) = (2 N / \lambda) {\rm tr}\,M^2 $\,.
In this case, the resolvent \eqref{R(z)} 
is given by 
\begin{equation}
R(z)_{\rm Gaussian}
=
\int_0^\infty dp {\rm e}^{-pz} f(p',N)\,, \qquad 
\bigg(p'= \sqrt{\lambda \over 4 N} p \bigg)\,.
\label{R(z)_Gauss}
\end{equation}
The function $f(p',N)$\,, which is defined in \eqref{W=Laguerre},
satisfies the following differential equation:
\begin{equation}
p' \partial_{p'}^2 f(p',N) 
+
3 \partial_{p'} f(p',N)
-
p'(4N + {p'}^2) f(p',N)
=
0\,. \label{Diff_eq}
\end{equation}
By using the relation $p'=\sqrt{\lambda/(4N)} p$
and by taking the large-$N$ limit with $p \ll N$\,,
\eqref{Diff_eq} is reduced to the 
Bessel differential equation and then we have 
\begin{equation}
f(p',N) \to {2 \over p \sqrt{\lambda}} I_1(p \sqrt{\lambda})\,,
\qquad 
(N\to \infty\,, \quad p\ll N)\,.
\label{modified-Bessel}
\end{equation}
Here $I_1$ is the modified Bessel function and the 
overall constant can be fixed by requiring $f(p',N) \to 1$
in the limit $p \to 0$.
Inserting this into \eqref{R(z)_Gauss}, and performing the
$p$-integral, we find the resolvent for the Gaussian potential:
\begin{equation}
R(z)_{\rm Gaussian}
= 
\int_0^\infty dp {\rm e}^{-pz}
{ 2 \over p \sqrt{\lambda}}
I_1(p \sqrt{\lambda})
=
{ 2 \over \lambda}
(z - \sqrt{z^2 - \lambda})\,.
\label{Wigner}
\end{equation}

Next we turn to the case with the insertion
of a Wilson loop with a large winding number $k$\ of ${\cal O}(N)$. 
In this case, 
we need to take account of effects of the 
Wilson loop on the saddle point, i.e., 
we should consider the following potential:
\begin{equation}
{\rm e}^{-V(M)}
=
\frac{1}{N} {\rm tr}\,{\rm e}^{kM}
{\rm e}^{-{2N \over \lambda} {\rm tr}\,M^2}\,.
\end{equation}
Hence the resolvent in the presence 
of the Wilson loop with winding number $k$, 
which we call $R_k(z)$, is defined by 
\begin{align}
& R_k (z) = \int_0^\infty dp \,{\rm e}^{-pz} \, 
W(p)_k\,, \label{Rk(z)}\\
& W(p)_k \equiv {1 \over Z_k} \int dM 
{1 \over N} 
{\rm tr}\,{\rm e}^{pM} 
\frac{1}{N}{\rm tr}\,{\rm e}^{kM} 
{\rm e}^{-{2N \over \lambda} {\rm tr}\,M^2 }\,, \label{W(k,p)} \\
&Z_k \equiv \int dM \frac{1}{N}{\rm tr}\,{\rm e}^{kM} 
{\rm e}^{-{2 N \over \lambda} {\rm tr}\,M^2}\,. \label{Zk}
\end{align}
Here we note that the computation in \eqref{Rk(z)} and \eqref{W(k,p)} 
should be done in the following order: 
first we calculate the 
``two loop correlator'' \eqref{W(k,p)} for finite $N$
and then we take the large-$N$ and large-$k$ limit 
with $\kappa$ and $p$ kept finite.
Finally we perform the Laplace transformation \eqref{Rk(z)}.

An explicit computation of the two loop correlator $W(p)_k$ 
is given in appendix \ref{app:2loop} by going to 
the eigenvalue integral and using the orthogonal polynomial method.
In the eigenvalue integral, we observe that 
the following decomposition,
\begin{equation}
{\rm tr}\,{\rm e}^{pM}\ {\rm tr}\,{\rm e}^{kM}
=
\sum_{i=1}^N \sum_{j \neq i} {\rm e}^{pm_i} {\rm e}^{km_j}
+
\sum_{i=1}^N {\rm e}^{(p+k) m_i}\,,
\label{trtr=ee}
\end{equation}
naturally arises and then these two terms are analyzed separately.
A physical interpretation of these terms will be discussed
shortly.
Here we just call the expectation value of the 
first term $w(k,p)$, and the second $w(k+p)$:
\begin{align}
W(p)_k = w(k,p) + w(k+p)\,. \label{W(p)kdecomp}
\end{align}
The explicit results of $w(k,p)$ and $w(k+p)$
are given in \eqref{w(k,p)final} and \eqref{w(k+p)final}:
\begin{align}
&
w(k,p)
=
f(p',N)
-
{1 \over N} {\rm e}^{{p'}^2 \over 2} 
\sum_{ij} {
{k'}^{i-j} L_{j-1}^{(i-j)} ( -{k'}^2 )
{p'}^{j-i} L_{i-1}^{(j-i)} ( -{p'}^2 )
\over
L_{N-1}^{(1)}(-{k'}^2)}\,, \label{w(k,p)main}\\
&
w(k+p)
=
{1 \over N} {f(k'+p',N) \over f(k',N)}\,. \label{w(k+p)main}
\end{align}

The first term on the right hand side of 
\eqref{w(k,p)main} gives the resolvent which
is identical to the one for 
the Gaussian potential \eqref{Wigner} in the large-$N$ limit 
after the Laplace transformation. 
The second term gives the ${\cal O}(N^{-1})$
resolvent \eqref{R(0,1)}, 
as shown in appendix \ref{sec:eval-secondterm}.
Here note that in the previous subsection, we assumed 
that
the ${\cal O}(N^{-1})$
resolvent \eqref{R(0,1)} has the cut
at $(-\sqrt{\lambda},\sqrt{\lambda})$ and has no
pole (see the second condition after \eqref{R(0,1)def}).
However in appendix \ref{sec:eval-secondterm}
we explicitly derive the resolvent \eqref{R(0,1)}
by taking the large-$N$, fixed $\kappa$ and $p$ 
limit for the second term in \eqref{w(k,p)main} 
and performing the Laplace transformation,
which justifies the assumption.

Next let us consider the Laplace transformation
of \eqref{w(k+p)main}\,.
We should take the large-$N$ limit with assuming the relation 
$p\ll N$ before the Laplace transformation.
Furthermore, this term depends on 
the additional parameter $k$ 
and the parameter region is chosen so that  
the combination 
$\kappa = k \sqrt{\lambda}/(4N)$ will be
kept finite.
The relevant terms in this limit are already given in \eqref{cal F}.
We just define $\tilde \kappa \equiv (k+p) \sqrt{\lambda}/(4N)$
and then \eqref{w(k+p)main} is evaluated in the
large-$N$ limit with $\kappa$ fixed and $p \ll N$ as 
\begin{align}
{1 \over N} {f(k'+p',N) \over f(k',N)}
& =
{1 \over N} 
\exp
\Big( 
- N 
\big(
{\cal F}_-(\tilde \kappa,N) 
- 
{\cal F}_- (\kappa,N)
\big)
\Big) \\
& =
{1 \over N} 
\exp
\Big(
p \sqrt{\lambda} \sqrt{1+\kappa^2}
+
{\cal O}(N^{-1})
\Big)\,.
\end{align}
By performing the Laplace transformation,
we obtain the following contribution to the resolvent of ${\cal O}(1/N)$:
\begin{equation}
R^{(2)}(z)
\equiv
\int_0^\infty dp {\rm e}^{-pz}
{1 \over N} 
{f(k'+p',N) \over f(k',N)}
\to
{1 \over N}
{1 \over z - z_\ast},
\quad
( z_\ast = \sqrt{\lambda} \sqrt{1+\kappa^2})\,.
\label{LaplacePole}
\end{equation}
This is exactly the pole corresponding to the isolated 
eigenvalue \eqref{isolated} found in the papers 
\cite{Yamaguchi:2006te,Yamaguchi:2007ps}.
Since the other ${\cal O}(N^{-1})$ contribution,
the second term in $w(k,p)$, is shown to give only the cut, 
these two ${\cal O}(N^{-1})$ contributions therefore give
clearly different contributions to the resolvent 
or eigenvalue distribution.

We have thus succeeded in identifying which correlator 
gives rise to the isolated pole, but, of course, 
the result \eqref{LaplacePole} itself is 
not quite new since we have just calculated
the position of the isolated eigenvalue by a different method. 
However, the above procedure implies the following 
interpretation of the Laplace transformation and the resolvent.

First note that, as we shall review in subsection \ref{DFreview},
the Wilson loop  
${\rm tr}\,{\rm e}^{kM}$ corresponds to the D3-brane with $k$ units
of the electric charge in the dual gravity picture 
in the AdS/CFT correspondence 
\eqref{RY-M}. Then the two loop correlator \eqref{W(k,p)} may 
be regarded as treating the D3-brane with string
charge $k$ as ``a part of the background'' and 
then introducing additional small number of string charge $p$\,,  
by which the total background is probed. 

Now we may advocate the following physical interpretation of each
term in the decomposition \eqref{trtr=ee}\footnote{%
In \cite{Yamaguchi:2007ps}, a similar decomposition of the exponential
operator is discussed in terms of world sheet nonperturbative
effects, in a slightly different context.};
the second term can be regarded as a situation where the additional
string charge $p$ is dissolved in the D3-brane with charge $k$, now
altering the number of the charge from $k$ to $k+p$.
On the other hand, the first term
may correspond to the case
in which the additional string charge $p$ is described by a 
probe D3-brane with charge $p$ that is separated from 
the ``background'' D3-brane with charge $k$, but has  
an interaction with it. 
At the leading order of the large-$N$ limit,
we may expect that the interaction between them 
is not taken into account and that the first term of
\eqref{w(k,p)main} describes two non-communicating D3-branes.
Actually
the first term in (\ref{w(k,p)main}) is blind to $k$ as a result of
cancellation of the effect from the D3-brane with charge $k$ between
the numerator and the denominator in \eqref{W(k,p)}, while the second
term in (\ref{w(k,p)main}) describes the interaction between these
D-branes in higher order of the string coupling constant.
It then seems natural that 
the first term of (\ref{W(p)kdecomp}) reproduces 
the semi-circle distribution,
which would carry information on the AdS$_5 \times$S$^5$, 
while the second term gives the pole, which is information
of the D3-brane with $k$ charge, 
in the leading contribution with respect to $1/N$ 
of each term. 

Based on the above derivation of $z_\ast$\,, we may also expect that,
in the gravity side, $z_\ast$ appears as an object which is conjugate 
to the string charge in some sense. 
We will further investigate these points from the gravity side 
in the following section.

\section{Matrix model resolvent from D3-brane picture}
\label{sec:D3-Wilsonloop} 

In this section we pursue the gravitational 
interpretation of the matrix model resolvent.
In particular we concentrate on the pole corresponding 
to the D3-brane with string charge $k$\,.
We will discuss the cut in the concluding section.

Since we found that the isolated 
pole is derived by the Laplace transformation of the expectation 
value of the operator ${\rm tr}\,{\rm e}^{(k+p)M}$,
it is rather easy to derive the pole itself in the 
D3-brane picture once the correspondence between the operator 
${\rm tr}\,{\rm e}^{kM}$ and the D3-brane solution with $k$ flux is established.
Indeed the expectation value of the operator 
${\rm tr}\,{\rm e}^{(k+p)M}$ in the large-$N$ 
and large-$k$ limit, or equivalently the function
${\cal F}(\tilde \kappa,N)$ is exactly reproduced 
by using the D3-brane solution as reviewed
in the following subsection\footnote{
More precisely, only the leading term ${\cal F}_0$,
i.e. the leading term of $\mathcal{F}_-$,
has been reproduced from the D3-brane solution.
Note that in the previous 
section, the pole has been derived by using 
the explicit form of ${\cal F}_0$ and also 
the fact that ${\cal F}_-$ can be expanded as
${\cal F}_- = {\cal F}_0(\kappa)+ (1/N) {\cal F}_1(\kappa) 
+ \cdots$, but the explicit form of higher terms,
including ${\cal F}_1$
are not needed.
}. 
So, what we need to do is just to expand it with respect to
$p$ and perform the Laplace transformation. 
This computation is just a repetition of the 
matrix model case.\footnote{
Here we should mention that this argument
does not fix the residue of the pole.
In order to do this, it is necessary to fix a normalization factor
in AdS/CFT correspondence up to the subleading order in $1/N$. 
There may exist another difficulty in determining a relative
normalization between the cut and the pole.
This is because, based on the viewpoint discussed at the end of the
previous section,
these two contributions correspond to different setups in the gravity
side, 
namely different perturbative vacua in string theory.}
However, as we pointed out in the previous section,
we should be able to extract more information 
about the role of the resolvent in the gravitational point of view.
In the rest of the paper, 
we will investigate this issue.

\subsection{A review of the Drukker-Fiol D3-brane solution}
\label{DFreview}
Let us give a brief review of the D3-brane solution
derived in \cite{Drukker:2005kx}, which will be 
useful for later discussions.
Their proposal is that the expectation value of the 
Wilson loop with winding number $k$ is given by 
the amplitude of a D3-brane carrying string charge 
$k$ and attached to the loop on the AdS boundary:
\begin{equation}
\bigg\langle {1 \over N} {\rm tr}\, {\rm e}^{kM} \bigg\rangle
=
\int_{\rm b.c.} 
{\rm e}^{-(S_{\rm D3}+S_{\rm b})}\,.
\label{AdS_conjecture}
\end{equation}
Here $S_{\rm D3}$ consists of the Dirac-Born-Infeld action $S_{\rm DBI}$
and the Wess-Zumino term $S_{\rm WZ}$ for a D3-brane:
\begin{align}
&
\label{SD3}
S_{\rm D3}
= S_{\rm DBI} + S_{\rm WZ}
=
\int d \sigma^1 d \sigma^2 d \sigma^3 d \sigma^4 {\cal L}\,,\\
&
S_{\rm DBI}
=
T_{\rm D3}
\int d \sigma^1 d \sigma^2 d \sigma^3 d \sigma^4 
\sqrt{ \det( g_{ab}+2 \pi \alpha' F_{ab} )}\,, \\
&
S_{\rm WZ}
=
- T_{\rm D3} \int {\cal P}[C_4]\,.
\end{align}
Here ${\cal P} [C_4]$ is the pullback of 
the four-form potential $C_4$
and the D3-brane tension is given by
$T_{\rm D3}=\frac{1}{(2\pi)^{3} l_s^4 g_s}$\,.
The path integral on the right hand side of \eqref{AdS_conjecture}
is taken over the D3-brane configurations
satisfying appropriate boundary conditions which 
are specified in terms of the Wilson loop.
The boundary term $S_{\rm b}$ is introduced in order 
to flip the boundary conditions for the world volume gauge field 
and the scalar fields.
In the semi-classical regime, 
the path integral 
may be well estimated by a saddle point value.

The authors of the paper started with 
the following metric for the AdS$_5$ part in the Poincar\'e coordinate:
\begin{equation}
ds^2 
= 
{L^2 \over y^2}
\big(
dy^2 + dr_1^2 + r_1^2 d \psi^2 + d r_2^2 + r_2^2 d\phi^2
\big)\,,
\end{equation}
with the curvature radius $L=(4 \pi g_s N)^{1/4}l_s =\lambda^{1/4}l_s$,
and considered a circle $r_1=R$, $r_2=0$ on the AdS boundary $y=0$
and then discussed the D3-brane solution which is attached to the circle.

The actual D3-brane solution and the evaluation 
of the action were discussed by using the following metric:
\begin{equation}
ds^2 
= 
{L^2 \over \sin^2 \eta}
\Big(
d \eta^2 
+ 
\cos^2 \eta d \psi^2 
+
d \rho^2 
+
\sinh^2 \rho
\big( d \theta^2 + \sin^2 \theta d \phi^2 \big)
\Big)\,.
\label{circlemetric}
\end{equation}
Here, $\eta$, $\rho$ and $\theta$ are related to $r_1$, $r_2$ and 
$y$ by the coordinate transformation
\begin{equation}
r_1 = { R \cos \eta \over \cosh \rho - \sinh \rho \cos \theta},\quad
r_2 = { R \sinh \rho \sin \theta \over \cosh \rho - \sinh \rho \cos
\theta }, \quad
y= {R \sin \eta \over \cosh \rho - \sinh \rho \cos \theta}\,.
\end{equation}
The range of these new variables are taken to be
$0 \leq \rho < \infty$, $0 \leq \theta \leq \pi$ and 
$0 \leq \eta \leq \pi/2$\,.
They chose $\psi$, $\rho$, $\theta$ and $\phi$
as the world volume coordinates by
$\sigma^1=\psi, \sigma^2=\rho, \sigma^3=\theta$ 
and $\sigma^4=\phi$
and assumed an 
S$^2$ symmetric and also $\psi$-translational invariant
ansatz, $\eta=\eta(\rho)$ 
and $F_{\psi \rho} = F_{\psi \rho}(\rho)$\,.
Then $S_{\rm DBI}$ and $S_{\rm WZ}$ are written down as
\begin{align}
S_{\rm DBI}
&
=
2 N \int d \rho d \theta
{ \sin \theta \sinh^2 \rho \over \sin^4 \eta}
\sqrt{ 
\cos^2 \eta(1 + {\eta'}^2)
+
( 2 \pi \alpha')^2 {\sin^4 \eta \over L^4}
F_{\psi \rho}^2
}\,,  \label{SDBI}\\
S_{\rm WZ}
&=
- 2 N \int d \rho d \theta
{\cos \eta \sin \theta \sinh^2 \rho \over \sin^4 \eta}
\bigg(
\cos \eta + \eta' \sin \eta 
{ 
\sinh \rho - \cosh \rho \cos \theta 
\over
\cosh \rho - \sinh \rho \cos \theta
}
\bigg)\,, \label{SWZ}
\end{align}
and the 
solution for the equations of motion is given by 
\begin{equation}
\sin \eta = \kappa^{-1} \sinh \rho, 
\quad
F_{\psi \rho} = {i k \lambda \over 8 \pi N \sinh^2 \rho}\,, 
\qquad 
\bigg(
\kappa = { k \sqrt{\lambda} \over 4 N}
\bigg)\,.
\label{DF_sol}
\end{equation}
Here $k$ is the string charge defined as\footnote{%
In this article we follow the notation of
\cite{Drukker:2005kx} for the D3-brane solution except this $\Pi$
which differs in the factor $i$.}
\begin{equation}
k = i \Pi \equiv
i \int d\theta d \phi 
{\partial {\cal L} \over \partial F_{\rho \psi}}\,,
\end{equation}
where $\mathcal{L}$ is defined as in (\ref{SD3}).
The solution is attached to the circle 
on the AdS boundary at $\rho = 0$\,.

The boundary terms are introduced in order to 
take account of the appropriate boundary conditions;
the D3-brane carries a fixed number of the string charge $k$,
and also the scalar fields transverse to the AdS horizon
should satisfy the Neumann boundary condition.
The latter boundary condition is motivated by the 
T-duality argument for the case with 
the fundamental string world sheet description instead of
the D3-brane \cite{Drukker:1999zq}\,.
The boundary condition for the gauge field 
is taken into account by adding the following boundary term:
\begin{equation}
S_A 
= \int d \psi d \theta d \phi 
{\partial {\cal L} \over \partial 
(\partial_\rho A_\psi)} 
A_\psi 
\bigg|_{\rho=0}\,.
\label{S_A}
\end{equation}
On the other hand, 
the relevant boundary term for the scalar fields 
is the one for the radial direction of AdS.
Here we need to take care of the fact that the boundary terms 
for the radial coordinate differ for different choices
of the coordinates system.
In \cite{Miwa:2006vd}, it was argued that 
a natural choice is to set the Neumann boundary condition
for the coordinate $u=1/y$:
\begin{equation}
S_u = \int d \psi d \theta d \phi 
{\partial {\cal L} \over \partial (\partial_\rho u)}u
\bigg|_{\rho=0}\,.
\label{S_u}
\end{equation}
One of the reasons for this choice was that 
the conjugate momentum of $u$ is finite at the AdS boundary, 
while that of $y$ is infinite. 
At the same time, this choice is necessary
for the correct cancellation of divergence.
In the following subsections we will see 
this choice is natural in our case as well. 

Summing up all the terms \eqref{SDBI}, \eqref{SWZ},
\eqref{S_A} and \eqref{S_u},
the classical value of $S_{\rm D3}+S_{\rm b}$
is evaluated as 
\begin{equation}
S_{\rm D3} + S_{\rm b} 
= 
- 2 N 
\big(
\kappa \sqrt{ 1 + \kappa^2}
+
{\rm arcsinh} \kappa
\big)\,,
\label{DFvalue}
\end{equation}
which indeed agrees with the matrix model result (\ref{cal F})
\cite{Drukker:2005kx}.
Therefore, as mentioned at the beginning of this section,
by replacing $\kappa$ with 
$\tilde \kappa = (k+p)\sqrt{\lambda}/(4N)$ and performing 
the Laplace transformation for ${\rm e}^{-(S_{\rm D3}+S_{\rm b})}$ 
with respect to $p$ ($p \ll N$)\,,
the pole at $z=z_\ast=\sqrt{\lambda}\sqrt{1+\kappa^2}$ 
can be reproduced.

In the above discussion, we used the explicit value
of the total action with the solution substituted,
which results in a function of $k$. 
As a result, the origin of the pole is not very clear.
Since the string charge carried by the solution should have been set 
by the boundary condition before solving the equations of motion,
 we should first change the boundary condition 
as $k \to k+p$ and then observe a response of the action.
It would allow us to discuss 
the Laplace transformation without using the 
explicit form of the solution from the beginning. 

Before doing this analysis, here it is useful to take much care of 
the boundary condition in the above analysis. 
In the above discussion, we imposed the boundary condition 
for the conjugate momentum of the gauge field as $i \Pi = k$\,.
On the other hand, we did not mention the condition 
on the boundary value of the conjugate momentum of $u$; 
it is just given by inserting the solution. 
In principle, the boundary condition for the 
conjugate momentum of $u$ should be specified independently before 
we discuss the solutions or the equations of motion. 
Moreover, based on the spirit of the 
AdS/CFT correspondence, the boundary condition should be 
specified in terms of the information contained 
in the Wilson loop.

In the case of the correspondence between the Wilson loops 
in fundamental representation and string world sheet, 
a detailed discussion on the boundary condition
was given in \cite{Drukker:1999zq}.
In the next subsection we consider a boundary condition 
for the D3-brane.  
This analysis plays an important role in subsection
\ref{LaplaceD3(2)}.

\subsection{D3-brane boundary condition imposed by Wilson loop}
\label{subsec:b.c.}

As we have mentioned, the boundary condition 
for the D3-brane should be specified in terms of 
the information carried by the Wilson loop. 
Before discussing the connection with the Wilson loop,
let us take a look at what boundary condition
is satisfied by the actual D3-brane solution \eqref{DF_sol}.
Near the AdS boundary, the radial coordinate $u$ can be approximated 
by $ u \sim 1/(R \eta)$. 
Hence the conjugate momentum of $u$, at the D3-brane boundary, 
is given by
\begin{equation}
{\partial {\cal L} \over \partial (\partial_\rho u)}\bigg|_{\rm \rho=0}
=
- R \eta^2 {\partial {\cal L} \over \partial (\partial_\rho
\eta)}\bigg|_{\rm \rho=0}
=
- {L^2 \over 2 \pi \alpha'} R k {\sin \theta \over 4\pi}\,.
\label{Pu}
\end{equation}
Here we introduce a new coordinate $U$ as 
\begin{equation}
U \equiv  
{L^2 \over 2 \pi \alpha'} u = {L^2 \over 2 \pi \alpha'} 
{1 \over y}\,.
\end{equation}
The coordinate $U$ is a natural choice for the scalar field, 
since $L^2/y$ corresponds to the radial coordinate 
of the flat six dimensions in the asymptotic region
of the original extremal black 3-brane solution.

We define $P_U$ as the conjugate momentum of $U$
integrated over S$^2$:
\begin{equation}
P_U 
\equiv 
\int d\theta d\phi 
{\partial {\cal L} \over \partial (\partial_\rho U)}\,,
\end{equation}
then the boundary condition is simply given by
\begin{equation}
\Pi = -i k\,,\quad P_U = - R k\,. \label{Pi-PU}
\end{equation}
Note that the factor $\sin \theta /(4\pi)$ in \eqref{Pu}
originates from the S$^2$ symmetric property of the solution.
Hence if we assume S$^2$ symmetry, 
we lose no information by integrating over the S$^2$.

In the case of string, the connection between 
the boundary value of the momentum and the information 
on the Wilson loop is proposed and its validity is supported by an argument 
using the Hamiltonian constraint in \cite{Drukker:1999zq}.
In our present case for the D3-brane, we consider the other way round:  
we examine what kind of conditions such a constraint implies 
on the boundary value of the momenta for the world volume fields, 
from which we can guess appropriate boundary conditions. 

For this purpose, first we note that 
the D3-brane solution in the previous subsection 
can be locally approximated by the solution 
corresponding to the straight Wilson line.
For later reference, we clarify this point. 
Let us consider a small region around a point on the loop,
say $\psi = \psi_0$ and $r_1=R$\,, 
and expand the coordinate as\footnote
{This corresponds to $\rho\rightarrow\epsilon\rho$, $\eta\rightarrow\epsilon\eta$.} 
\begin{equation}
\psi = \psi_0 + \epsilon t\,, \quad
r_1 = R(1 + \epsilon \rho \cos \theta)\,, \quad
r_2 = R \epsilon \rho \sin \theta\,, \quad
y = R \epsilon \eta\,.
\label{circle to straight}
\end{equation}
Here $\epsilon$ is a small parameter and we 
assume that the coordinates $t$\,, $\rho$\,, and $\eta$ 
take values of order $1$\,.
The angular coordinate $\theta$ can take value
in the original range as $0 \leq \theta \leq \pi$\,.
Then the AdS$_5 \times$S$^5$ metric is reduced to the following one:
\begin{equation}
ds^2
= 
{L^2 \over \eta^2} 
(
d\eta^2 
+ 
dt^2 
+ 
d\rho^2 
+ 
\rho^2 (d \theta^2 + \sin^2 \theta d \phi^2))
+
L^2 d \Omega_5^2,
\label{localAdS}
\end{equation}
and the Wilson loop is extended along the $t$-direction. 
The solution \eqref{DF_sol} becomes the one 
corresponding to the straight Wilson line in $\epsilon\rightarrow 0$ limit. 
In the following we discuss a kind of Hamiltonian constraint
only for the case of the straight line. 
However, from this observation, we expect that the analysis can also
be applied for
a local patch of a Wilson loop of an arbitrary (smooth) shape. 
Here we note that as we will see, when we interpret our analysis 
as the local version of the analysis for a circle, 
we must take the scale factor $R$ and $\epsilon$ into account, 
which is lost in the case of the straight line. 

Moreover, it should be stressed that in the following argument,
although we use the S$^2$ symmetric ansatz,
we do not use information on the classical solution reviewed in the
previous subsection (nor its reduction to that for the straight line)
 in the gravity side.
In our semi-classical analysis, it seems natural to 
allow any field fluctuation around a classical configuration
on the right hand side in \eqref{AdS_conjecture} 
as long as it is consistent with the boundary condition.
From such a point of view, the S$^2$ symmetric ansatz
is just an assumption for allowed fluctuations.\footnote{
It may also be possible that, as a definition of the correspondence, 
the functional integral on the right hand side of 
\eqref{AdS_conjecture} should be constrained 
in a domain where the world volume fields preserve some  
symmetries of the Wilson loop. 
In such a case, the S$^2$ symmetric ansatz becomes more important 
than in the case where it is just an assumption.
Although it is an important problem to give a precise definition 
for the integration region on the right hand side of 
\eqref{AdS_conjecture}, it is beyond the scope of the present article.} 
Of course requiring global symmetries including the supersymmetries 
is sometimes so strong 
that an allowed configuration will be uniquely a classical solution,
especially when the Wilson loop is highly symmetric 
as for the straight line and the circular loop;
see appendix \ref{App:BPS-cond}.
However here we emphasize that the S$^2$
symmetric ansatz still allows a wide range of fluctuations, 
and the following argument should hold for such field fluctuations. 

We use the following coordinates for 
the AdS$_5 \times$S$^5$ geometry:
\begin{align}
ds^2 
&=
\bigg(
{2 \pi \alpha' \over L}
\bigg)^2 U^2 
\big( dt^2 + d\rho^2 + \rho^2 ( d \theta^2 + \sin^2 \theta d\phi^2) \big)
+
L^2 { (dU^i)^2 \over U^2} \\
&=
G_{tt} dt^2 + G_{\rho\rho} d\rho^2 + G_{\theta \theta} 
( d \theta^2 + \sin^2 \theta d \phi^2) + G_{U^i U^i} (dU^i)^2 \\
&=
G_{X^M X^N} dX^M dX^N
\,. 
\end{align}
The four directions $(t,\rho,\theta,\phi)$ are
identified as the four-dimensional space where 
the gauge theory lives. The Wilson loop is 
located at $\rho=0$\, and extended into $t$-direction.
We are now discussing the D3-brane configuration which 
is pinched to the line $\rho=0$ on the AdS boundary, $U=\infty$\,.
We introduce the world volume coordinate $\sigma^a$ ($a= 1 \sim 4$)
and, as mentioned above, assume the S$^2$ symmetric ansatz\footnote{%
There are several S$^2$ in this AdS$_5 \times$ S$^5$ geometry to be
identified with the world volume S$^2$,
for example, some S$^2 \subset$ S$^5$. 
Here we consider configurations which are trivial in the S$^5$
direction.
This is also an assumption.}:
\begin{align}
&\theta =\sigma^3\,, \quad \phi=\sigma^4\,, \\
&t=t(\sigma^1,\sigma^2)\,, \quad \rho=\rho(\sigma^1,\sigma^2)\,, \quad
U^i=U^i(\sigma^1,\sigma^2)\,, \quad F_{12}=F_{12}(\sigma^1,\sigma^2)\,.
\end{align}
We take the coordinate $\sigma^2$ in such a way that
the D3-brane boundary is at $\sigma^2=0$, i.e.,
$\rho(\sigma^1,\sigma^2=0)=0$ and $U(\sigma^1,\sigma^2=0)=\infty$.
Then on the boundary $\sigma^2=0$, 
the parameter $\sigma^1$ should parameterize
 the loop on which the D3-brane is attached.

Now the D3-brane action is given by
\begin{align}
&
S_{\rm DBI} + S_{\rm WZ}
=
\int d \sigma^1 d \sigma^2 L_{\rm D3} 
=
\int d \sigma^1 d \sigma^2 (L_{DBI}+L_{WZ})\,, 
\label{action}\\
&
L_{\rm DBI}
=
4 \pi T_{\rm D3} 
G_{\theta \theta}
\sqrt{g_{11}g_{22}-g_{12}^2 + (2 \pi \alpha')^2 F_{12}^2}\,,\\
&
L_{\rm WZ}
=
- 4 \pi T_{\rm D3} 
G_{\theta \theta}
\sqrt{G_{tt}G_{\rho\rho}} 
(\partial_1 t \partial_2 \rho 
- \partial_1 \rho \partial_2 t)\,.
\end{align}
Here the Lagrangian $L_{\rm D3}$ and $L_{\rm DBI,WZ}$ are
defined with the S$^2$ part integrated.

We define the conjugate momentum $P_{X^M}$ and $\Pi^a$ 
for $X^M$ and $A_a$, respectively as
\begin{equation}
P_{X^M}
=
{\partial L_{\rm D3} \over \partial (\partial_2 X^M)}\,, \quad
\Pi^a
=
{{\partial L_{\rm D3}} \over \partial ({\partial_2 A_a})}\,.
\label{conj-mom}
\end{equation}
Then, the following identity holds:
\begin{equation}
G^{tt}(P_t - {\cal P}_t)^2
+
G^{\rho\rho}(P_\rho - {\cal P}_\rho)^2
+
G^{U^i U^i}(P_{U^i})^2
+
{1 \over (2 \pi \alpha')^2}
(\Pi^1)^2 g_{11}
=
(4 \pi T_{\rm D3} G_{\theta \theta})^2 g_{11}, 
\label{constraint1}\\
\end{equation}
where 
$ g_{11}=G_{tt} (\partial_1 t)^2
+G_{\rho\rho} (\partial_1 \rho)^2+G_{U^i U^i}(\partial_1 U^i)^2$.
The ``momenta'' ${\cal P}_{X^M}$ are defined as 
${\cal P}_{X^M} \equiv \partial L_{\rm WZ} 
/ \partial (\partial_2 X^M)$\,,
whose explicit forms are given by 
\begin{equation}
{\cal P}_t = 4 \pi T_{\rm D3} G_{\theta \theta} 
\sqrt{G_{tt}G_{\rho\rho}}\partial_1\rho, 
\quad
{\cal P}_\rho = - 4 \pi T_{\rm D3} G_{\theta \theta}
\sqrt{G_{tt}G_{\rho\rho}}\partial_1 t\,. \label{constraint2}
\end{equation}
These do not include any derivative 
with respect to $\sigma^2$ and thus \eqref{constraint1}
can be regarded as a constraint in the phase space.
By using the explicit form of the metric,
the constraint \eqref{constraint1} can be rewritten as
\begin{align}
0&
=
(P_{U^i})^2 
+ 
(\Pi^1)^2 \big( (\partial_1 t)^2 + (\partial_1 \rho)^2 \big) 
-{16 \pi \over \lambda} N 
\big(P_t \partial_1 \rho - P_\rho \partial_1 t \big) \rho^2 \notag \\
& \quad 
-\bigg(
{64 \pi^2 \lambda^{-2}} N^2 (U \rho)^4 
- 
{\lambda \over 4 \pi^2} (\Pi^1)^2
\bigg)
\bigg(
{\partial_1 U^i \over U^2}
\bigg)^2 
+
{\lambda \over 4 \pi^2 } (P_t^2 + P_\rho^2) {1 \over U^4}.
\end{align}

Let us assume that the D3-brane configuration 
asymptotically satisfies the following conditions
in the limit $\sigma^2 \to 0$:
\begin{equation}
U \to \infty\,, \quad
U \rho < \infty\,, \quad
P_t,\, P_\rho \, 
\lesssim \, 
P_U,\, \Pi^1 
\, (\sim \, N) \,.
\label{conditions}
\end{equation}
The first condition has already been imposed as a boundary condition,
which requires that the D3-brane 
should be attached to the AdS boundary.
The second means that the radius of the
S$^2$ part of the D3-brane is finite at the boundary.
The last one refers to the behavior of the fields
at the boundary with respect to $N$.
Although the last two inequalities are new assumptions, 
we expect that \eqref{conditions} picks up a reasonable set of fluctuations.

Under the conditions \eqref{conditions}, 
the equation \eqref{constraint1}, 
in the limit $\sigma^2 \to 0$,
is reduced to a simple form:
\begin{equation}
(P_{U^i})^2 
+ 
(\Pi^1)^2 (\partial_1 t)^2 =0\,.
\label{Pi2+P2=0}
\end{equation}
Thus we find that the world volume fields on the D-brane 
attached to the Wilson loop at the AdS boundary are subject to 
this constraint, at least under the S$^2$ symmetric ansatz. 
Now it is natural to impose the boundary condition
on $\Pi^1$  as  
\begin{align}
\Pi^1(\sigma^1,\sigma^2=0)=-ik,  \qquad \mbox{for each~} \sigma^1,
\end{align}  
since the boundary condition imposed by the straight 
Wilson line, with constant $\theta^i$\,,
should not depend on the parameter $\sigma^1$\,.
Then since the conjugate momentum of $U^i$ and that of $U$\,,
which are defined by \eqref{conj-mom}, 
are related through $P_{U^i} = (U^i/U) P_U $
on the boundary,
the equation \eqref{Pi2+P2=0} tells us that  at $\sigma^2=0$, 
\begin{equation}
\Pi^1= -i k\,, \quad 
P_{U^i} = P_U \Theta_i = - k | \partial_1 t | \Theta_i\,,\quad 
\Theta_i \equiv U^i/U\,. 
\label{Pi-PU(2)}
\end{equation}
Note that there should be a sign ambiguity for $P_U$.
Here we took the minus sign so that it will be consistent with the 
classical solution. 
This is the only input from the classical solution.

As an application, in order to reinterpret \eqref{Pi-PU(2)}
in the case of the circular loop, we need to 
take account of the scale factor $R$ and $\epsilon$
as pointed out in the discussion after \eqref{localAdS}.
Since the first equation of \eqref{Pi-PU(2)} 
dose not have any uncontracted target indices,
it does not need to be changed.
On the other hand, the second equation should 
be reinterpreted by taking account of the scale
factor as 
$ U ({\rm straight}) = R \epsilon \times U ({\rm circle}) $ 
as implied by the last equation in \eqref{circle to straight}, 
and $ t = \epsilon^{-1} \psi$\,.
Hence we have
\begin{equation}
P_U = - k |\partial_1 t| \,\, (\textrm{for straight line})
\quad \to \quad
P_U = - k R |\partial_1 \psi| \,\, (\textrm{for circular loop}).
\label{straight->circle}
\end{equation}
This is precisely the relation 
we found at \eqref{Pi-PU}
(note that the \eqref{straight->circle}
dose not depend on the choice of the parameter
$\sigma^1$)\,.

So far we have considered the local straight line \eqref{Pi-PU(2)} 
or the circular loop \eqref{straight->circle}.
In more general case, we thus expect  
following boundary conditions:
\begin{equation}
\Pi^1 = - i k\,, \quad
P_{U^i} = - k |\partial_1 X| \Theta_i\,.
\label{Pi-PU(3)}
\end{equation}
Here $X^\mu$ is the four-dimensional Cartesian coordinate, 
i.e., $ds_{{\rm AdS}_5}^2 = Y^{-2}( dY^2 + dX^\mu dX^\mu)$\,
which is identified with the space where the gauge theory lives.

In the spirit of the AdS/CFT correspondence,
the boundary condition should be 
given in terms of the Wilson loop:
\begin{equation}
\label{Wkingauge-theory}
W_k(C) 
= 
{\rm tr}\,{\rm P}\exp
\bigg(
\int ds 
\Big(
i A_\mu \dot x^\mu + \Phi_i |\dot x| \theta^i 
\Big)
\bigg)\,.
\end{equation}
Thus we find that 
natural boundary conditions are  
\begin{align}
&
X^\mu(\sigma^1=\sigma^1(s),\sigma^2=0
)=x^\mu(s)\,, 
\label{D} \\
&\Pi^1(\sigma^1,\sigma^2=0) = - i k\,, 
\quad 
P_{U^i}(\sigma^1=\sigma^1(s),\sigma^2=0) = - k  |\dot x(s)| \theta_i (s)\,,
\label{N}
\end{align}
where we omitted the trivial S$^2$-dependence in \eqref{D}
and $\sigma^1(s)$ determines a relation between two
parameters of the loop (or the boundary of the world volume) 
representing the same point. Hereafter we assume $\sigma^1=s$ 
by using the reparametrization invariance of the Wilson loop. 
We find that the Wilson loop gives the Dirichlet boundary conditions 
\eqref{D} on the scalar fields along the world volume direction 
of the $N$ D3-branes and the Neumann boundary conditions 
\eqref{N} on the gauge field and the scalar fields perpendicular 
to the world volume direction, as expected \cite{Drukker:2005kx}.  

We would make some remarks. First we notice that in \eqref{N} 
the integration over S$^2$ in AdS$_5$ is done in $\Pi^1$ and $P_{U^i}$, 
which is seen from the definitions \eqref{action} and \eqref{conj-mom}. 
We expect that this integration is trivial even 
for loops of arbitrary shape 
if the loop is locally approximated by a straight line
with the S$^2$ symmetric ansatz.
Next, we note that our boundary conditions are natural generalization of 
the ones given in \cite{Drukker:1999zq} in the sense that 
\eqref{D} and the second boundary condition in \eqref{N} amounts to them 
in the case of the string world sheet, namely when we neglect the
$\sigma^3$, $\sigma^4$-dependence and use the Nambu-Goto action. 

Finally let us make a remark on the most important aspect of  our
boundary conditions.
By use of the embedding coordinates $X^\mu$, we can convert 
the world volume indices of $\Pi^a$ into the space-time indices
along the world volume like $\Pi^\mu \equiv \partial_a X^\mu \Pi^a$\,.
Since $\Pi^a$ has the only non-vanishing component for $a=1$,
$\Pi^\mu$ satisfies the following boundary condition 
\begin{equation}
\Pi^\mu =
\partial_1 X^\mu \Pi^1
= -ik \dot{x}^\mu \,,\qquad \mbox{at~} \sigma^2=0,
\end{equation}
where we have used \eqref{D}.
Using the second boundary condition in \eqref{N}, this gives the
following relation
\begin{equation}
  \label{eq:CM-likerelation}
 ( \Pi^\mu)^2 + (P_{U^i})^2 =0 \,. 
\end{equation}
It is worth noting that the boundary condition 
\eqref{N} imposed by the Wilson loop \eqref{Wkingauge-theory} 
thus corresponds to the BPS condition in 
\cite{Callan:1997kz}, i.e., force balance between 
the electric charge $\Pi^1$ and the deformation of 
the D3-brane which is characterized by $P_U$\,,
in the case of the spike solution in the flat space.
The force balance equation becomes simple and symmetric form
for our choice of the coordinate, i.e., $X^\mu$ and $U^i$\,.
This may be understood by the fact that the coordinate system
$\{X^\mu, X^{i+4}=2 \pi \alpha' U^i \}$ corresponds to the
Cartesian coordinates of the flat ten dimensional space 
in the asymptotic region of the black 3-brane solution.
Thus it is natural that the equation (\ref{eq:CM-likerelation}) 
also implies a local BPS condition for the Wilson loop. 
In fact, in gauge theory side, we can introduce
a loop of an arbitrary shape 
in the internal space by replacing $|\dot x|\theta^i$ in 
\eqref{Wkingauge-theory} with some function $\dot y^i(s)$.
For a corresponding D3-brane solution our boundary condition 
becomes the following symmetric form: 
\begin{equation}
\Pi^{\mu}(\sigma^1,\sigma^2=0)=-ik\dot x^\mu(\sigma^1),  \qquad
P_{U^i}(\sigma^1,\sigma^2=0) = - k\dot y_i(\sigma^1)\,,
\label{symmetricbc}
\end{equation}
and then the force balance condition is equivalent to the relation
\begin{equation}
  \dot x^2 = \dot y^2 \, ,
\end{equation}
which is nothing but the local BPS condition for Wilson loops 
in the gauge theory side\cite{Drukker:1999zq,Zarembo:2002an}.
Therefore we find our boundary condition quite natural 
because once it is assumed, 
the local BPS conditions in both sides become equivalent\footnote{%
However this equivalence may be a peculiar feature of AdS$_5 \times$
S$^5$ background\cite{Chu:2008xg}\,.}. 

Another interesting point is that the Gauss law constraint 
implies that the coordinates
$\sigma^1$ and $\sigma^2$, i.e. 
normal direction and tangential direction of the loop,
are mutually independent.
Indeed, from \eqref{N} and \eqref{symmetricbc} 
\begin{align}
0=\Pi^{a=2}=\frac{\partial \sigma^2}{\partial X^\mu}(-ik\dot x^\mu), \qquad
-ik\dot x^\mu=\Pi^\mu=\frac{\partial X^\mu}{\partial \sigma^1}\Pi^1
                    =\frac{\partial X^\mu}{\partial \sigma^1}(-ik), 
\end{align} 
and hence $\partial\sigma^2/\partial\sigma^1=0$. 
We would also like to note that \eqref{symmetricbc} makes 
the boundary terms (\ref{S_A}) and (\ref{S_u}) 
manifestly invariant under the reparametrization.\footnote
{It might be interesting to notice that the boundary term 
(\ref{S_A}) and (\ref{S_u}) takes the form of the exponent of
the $U(1)$ Wilson loop 
under our boundary conditions, 
though we should not be confused about the gauge field 
in the super Yang-Mills theory and that on the D-brane world volume.}

\subsection{Laplace transformation and matrix model resolvent revisited}
\label{LaplaceD3(2)}
Let us reconsider the Laplace transformation of the
circular Wilson loop taking account of 
the following boundary condition
\begin{equation}
\Pi = -ik, \quad P_U = - R k\,. \label{b.c.}
\end{equation}
We now evaluate the right hand side of \eqref{AdS_conjecture} 
under this boundary condition.
In a region like $\lambda\gg 1$ the saddle point approximation will be
valid, and then
 we can estimate 
the summation $S_{\rm D3} + S_{\rm b}$ by its saddle point value.
Since adding the boundary terms amounts to the Legendre
transformation, this saddle point value is
given by a function of the boundary values of momenta as
$(S_{\rm D3} + S_{\rm b})|_{\rm saddle \ point} 
= {\cal R}(\Pi,P_U)$\,, where $\Pi$ and $P_U$ express the 
boundary values of the momenta.\footnote{
Here we still assume the S$^2$ symmetric ansatz.
Hence only the conjugate momentum of 
$\eta(\rho)$ and $F_{\rho \psi}(\rho)$ are considered.}
Then the AdS/CFT correspondence for the Wilson loop 
in a semi-classical limit claims 
\begin{equation}
\bigg\langle
{1 \over N} {\rm tr}\,{\rm e}^{kM}
\bigg\rangle
=
{\rm e}^{-{\cal R}(-ik,-Rk)}\,.
\end{equation}
As mentioned at the beginning of this section, 
in order to derive the matrix model resolvent 
associated with the isolated eigenvalue,
we change the boundary condition as $k \to k+p$ 
and make the Laplace transformation with respect to $p$\,,
with taking account of the relation $p \ll N$\,. 
Now let us follow simply this procedure 
without using other information like 
the explicit form of the saddle point configuration 
in contrast to the derivation described below \eqref{DFvalue}. 
In fact, this approach enables us to clarify a gravitational 
interpretation of the isolated eigenvalue as we will see shortly.   

Recalling that
the variation of ${\cal R}$ with respect to change of the boundary value 
originates only from that of the boundary terms 
\cite{Drukker:2005kx} as\footnote
{It is important to notice that this is true because of the saddle
  point equation. 
Hence this is not generally true for general configurations.
}
\begin{equation}
\delta {\cal R}(\Pi,P_U) 
= 
\int d \psi \delta (\Pi) A_\psi\bigg|_{\rho = 0}
+
\int d \psi \delta (P_U) U \bigg|_{\rho=0}.
\label{R-variation}
\end{equation}
By neglecting the terms which vanish in the limit $p \ll N$, 
${\cal R}(-i(k+p), -R(k+p))$ 
is rewritten as
\begin{equation}
{\cal R}(-ik,-Rk)
-
p 
\bigg(
\int d \psi ( i A_\psi + R U)
\bigg)
\,.
\label{R-expand}
\end{equation}
Then by performing the Laplace transformation,
we can derive the following pole term:
\begin{equation}
\int_0^\infty dp {\rm e}^{-pz} {\rm e}^{-{\cal R}(-i(k+p),-R(k+p))} \, 
\to \,
{ {\rm e}^{-{\cal R}(-ik,-Rk)} \over z- \int d \psi (i A_\psi + R U)}\,.
\label{ochi}
\end{equation}
As a check, we shall plug the solution \eqref{DF_sol} into this, 
and then the location of the pole reproduces
the position of the isolated eigenvalue correctly:
\begin{equation}
\int d\psi (i A_\psi +  R U)
=
\sqrt{\lambda} \sqrt{1 + \kappa^2}\,.
\label{conclusion}
\end{equation}

As we have noted a couple of times in this paper, it is not so
striking a result in itself that the position of the isolated eigenvalue is
reproduced.
Here we claim that the position of the pole, 
namely the position of the eigenvalue, 
has been identified with the electric flux $F_{\rho \psi}$ 
integrated with respect to $\rho$ and $\psi$ directions 
of the D3-brane (plus contribution from scalar field $U$) 
as far as the isolated eigenvalue is concerned.
This is exactly what we anticipated 
in the last of section \ref{sec:WLGT}, 
because this value is conjugate to the electric flux,
namely the string charge, 
in the sense of \eqref{R-variation} and \eqref{R-expand}. 

Here it is important to recognize that in deriving the gravitational 
interpretation of the eigenvalue \eqref{conclusion}, our boundary condition 
\eqref{b.c.} plays a crucial role. 
This is because each boundary term \eqref{S_A} and \eqref{S_u}
diverges at the boundary, but after an appropriate regularization,
dependence on the regularization cancels out between them as shown
in \cite{Drukker:2005kx}. One can easily find that under our boundary
condition \eqref{b.c.}, the divergence again cancels because of the fact that 
both of boundary values of $\Pi$ and $P_U$ are given in terms of the same $k$. 
In particular,  if we impose the boundary condition 
only on $\Pi$, this is not the case.  
Finally, it is interesting to see that 
from \eqref{conclusion} the gravitational counterpart
of the isolated eigenvalue 
takes the form of the exponent of the $U(1)$ Wilson loop 
(see the footnote at the end of subsection \ref{subsec:b.c.}).

\section{Conclusions and discussions}
\label{sec:conclusion}

In this paper, we have analyzed the gauge theory Wilson loop
winding around a circular loop $k$ times, by using a D3-brane carrying
$k$ units of string charge 
in the context of AdS/CFT correspondence.
It is known that the calculation of the expectation value
of this Wilson loop, thanks to its symmetry, boils down 
to considering a Gaussian matrix model with
an exponential operator insertion.
We have then aimed our goal at establishing a
gravitational interpretation of the eigenvalue in this matrix model
based on the D3-brane version of the correspondence
\eqref{RY-M}.
This point presents a contrast to the 
preceding papers on ``bubbling Wilson loops''
\cite{Lin:2004nb,Yamaguchi:2006te,Lunin:2006xr}.

We started with analyzing the resolvent of the Gaussian matrix model 
using the orthogonal polynomials. 
In the calculation, 
we first derive the resolvent 
in its inverse Laplace transformed form, by use of ${\rm e}^{pz}$\,.
Then after the Laplace transformation we obtained the usual resolvent.
In the inverse Laplace transformed form, 
we derived the expression which is valid to all orders in $1/N$\,, 
and we identified a term that corresponds 
to the eigenvalue which is isolated 
due to a large $k \sim \mathcal{O}(N)$ effect. 
The remaining terms are 
responsible for the rest of the resolvent.
The part corresponding to the isolated eigenvalue has the structure
in which $p$ is included in the shift of $k$ as 
${\rm tr}\, {\rm e}^{kM} \to {\rm tr}\, {\rm e}^{(k+p)M}$\,.
In accordance with the D3-brane description of 
the matrix model operator ${1\over N}{\rm tr}\, {\rm e}^{k M}$, 
we were naturally led to consider the D3-brane 
with $k+p$ string charge and the Laplace transformation of its amplitude. 
Eventually we identify the position of the isolated eigenvalue, 
observed as an isolated pole of the resolvent, 
with an integrated flux
 (and scalar fluctuation)
 on the D3-brane.
We therefore succeed in providing, at least in part,
a gravitational interpretation of the eigenvalue in the Gaussian
matrix model.

As a by-product, we have proposed natural boundary conditions 
for the D3-brane configuration with fluxes in terms of the Wilson loop.
These boundary conditions also provide a direct relationship 
between local BPS conditions in the gauge theory side and that of the
effective theory on the probe D3-brane.

Let us now discuss the cut of the resolvent.
In the present paper, we have mainly studied the 
gravitational description of the isolated pole of the resolvent.
It is surely nice if the cut of the resolvent 
can also be discussed based on our gravitational 
description. 
In the gauge theory side we saw in subsection \ref{sec:Laplace}
that the cut of the leading semi-circle originates from the leading
term in $w(k,p)$.
In the gravity side, we argued that this term 
corresponds to the configuration with a D3-brane 
carrying string charge $p\,(\ll N)$ in addition to the 
one with the string charge $k\,(\sim N)$ which is now 
treated as a part of the background.
So, we may expect that by performing the
Laplace transformation of the contribution from such
configuration, the cut would be reproduced.
However, here we should recall that the saddle point value 
of the gravity side gives only the leading term 
of the gauge theory observables.
In fact, the saddle point value of the D3-brane action 
with string charge $p\,(\ll N)$ naively gives 
the result ${\rm e}^{p\sqrt{\lambda}}$\,.
This is just the leading contribution
of the Bessel function \eqref{modified-Bessel}
in the large 't Hooft coupling limit, and it dose not
lead to the cut after the Laplace transformation.
Hence, it is clear that in order to reproduce 
the cut, we need to take account of the quantum $\alpha'$ 
correction of the D3-brane with string charge $p$.

Another point we should also mention is the case of the 
correspondence between anti-symmetric Wilson loop
and D5-brane \cite{Gomis:2006sb,anti-sym-WilsonLoops}.
It would be interesting to examine 
whether our gravitational interpretation 
is valid also in this case.
However the situation is not straightforward.
This is because the multiplication of 
the $k$-th anti-symmetric trace operator ${\rm Tr}_{A_k} {\rm e}^M $ 
and the probe operator ${\rm tr}\, {\rm e}^{pM}$ dose not include the 
($k+p$)-th anti-symmetric trace operator.
It seems to suggest
that we need to consider the configuration with 
a D3-brane carrying $p$ flux
around the background D5-brane with $k$ flux, 
instead of only considering a single D5-brane with ($k+p$)-flux.

Here we also point out a subtlety concerning the relation between a
Wilson loop with winding number $k$ 
and that in the $k$-th symmetric representation.
Let $u_i$ be $i$-th eigenvalue of ${\rm e}^M$ where $M$ is the matrix
variable used in the matrix model analysis in section \ref{sec:Laplace}.
The Wilson loop with winding number $k$ can be calculated by
 ${\rm tr}\, {\rm e}^{k M} = \sum_i u_i^k$
 while the Wilson loop in the $k$-th symmetric representation
corresponds to
\begin{equation}
\label{eq:symmetric-loop}
  \text{Tr}_{S_k} \, {\rm e}^M
 \equiv \sum_i u_i^k + \sum_i \sum_{j \neq i} u_i^{k-1} u_j
+ \cdots \,,
\end{equation}
where $\cdots$ includes the other combinations of the powers of the
eigenvalues.
See \cite{Gomis:2006sb,Yamaguchi:2007ps} for the details.
Here we note that the decomposition that appeared in our matrix model
calculation (\ref{W(p)kdecomp}) is equivalent to the terms here when
we write them as $w((k-1)+1) + w(k-1,1)$ with $k$ being
$k-1$ and $p$ being $1$.
It has been argued that the expectation values of these
two Wilson loops
coincide under the large 't Hooft coupling limit, namely in
(\ref{eq:symmetric-loop}), compared to the first term,
the rests are exponentially suppressed in large $\lambda$.
However our explicit calculation shows that $w(k)$ is of order
$1/N$ compared to $w(k-1,1)$.
Note that $k$ and $p$ have been assumed to be of $\mathcal{O}(N)$ and
$\mathcal{O}(1)$ respectively, and then the calculation is still valid.
The equivalence thus holds only when the large-$\lambda$ limit
overcomes the difference in $1/N$, that is, $\lambda$ needs to be
larger than $\log N$.
This is therefore out of the usual limit such as first taking 
$N \rightarrow \infty$ with fixed $\lambda$ and then taking $\lambda$ to
be large.
Note that in this article we employ only the fact that in the strong
coupling limit the D3-brane solution with $k$-flux agrees with
${\rm tr}\, {\rm e}^{kM}$,
and then this subtlety does not matter.

Finally, it would be also important and interesting future work to 
clarify the relation between our viewpoint and the bubbling picture
\cite{Yamaguchi:2006te,Lunin:2006xr}.
We believe our viewpoint leads to a deep understanding 
of the gravitational interpretation of the eigenvalue, 
or moreover the connection between gauge theory and gravity.

\section*{Acknowledgements}
The authors would like to thank S. Yahikozawa, 
S. Yamaguchi, T. Yoneya.
They also thank Dimitrios Giataganas and Nadav Drukker for helpful
comments on the preprint version of the article. 
The work of A. M. is supported in part by JSPS 
Research Fellowships for Young Scientists.
We thank the Yukawa Institute for Theoretical Physics 
at Kyoto University. 
Discussions during the YITP workshop YITP-W-08-04 on 
``Development of Quantum Field Theory and String Theory'' 
were useful to complete this work.

\appendix 

\section{Two loop correlator in Gaussian matrix model}
\label{app:2loop}
In this appendix we compute the ``two loop correlator''
of the Wilson loop:
\begin{align}
&W(p)_k 
= {1 \over Z_k}
\int d M 
{1 \over N} {\rm tr}\, {\rm e}^{pM}
{1 \over N} {\rm tr}\, {\rm e}^{kM}
{\rm e}^{-{2 N \over \lambda} {\rm tr}\,M^2}\,,
\label{twoW} 
\\
&
Z_k \equiv
\int dM 
{1 \over N} {\rm tr}\,{\rm e}^{kM} 
{\rm e}^{-{2N \over \lambda} {\rm tr}\,M^2}\,,
\end{align}
in order to examine the resolvent in the presence of the Wilson loop 
with a large winding number. 
For this purpose, we have to evaluate \eqref{twoW} for finite $N$ 
as noticed below \eqref{Zk}. Therefore, we calculate \eqref{twoW} 
by means of the orthogonal polynomials. 
We first change the variables as
\begin{equation}
k' = \sqrt{ \lambda \over 4 N} k\,, \quad
p' = \sqrt{ \lambda \over 4 N} p\,, \quad
M' = \sqrt{ 4N \over \lambda } M\,,
\end{equation}
then $W(p)_k$ becomes
\begin{align}
& W(p)_k= 
{1 \over \tilde Z_k}
\int dM' 
{1 \over N} 
{\rm tr}\,{\rm e}^{p' M'}
{1 \over N} 
{\rm tr}\,{\rm e}^{k' M'}
{\rm e}^{-{1 \over 2}{\rm tr}\,{M'}^2}\,, \\
&\widetilde Z_k  
= 
\bigg( { 4 N \over \lambda } \bigg)^{{N^2 \over 2}} Z_k 
=
\int d M' {1 \over N} {\rm tr}\,{\rm e}^{k' M'}
{\rm e}^{-{1 \over 2}{\rm tr}\,{M'}^2}\,.
\end{align}
Using the Hermite polynomial $P_i(m)$ satisfying 
\begin{equation}
\int dm \, {\rm e}^{-{1\over 2}m^2}
P_i(m) P_j(m) = h_i \delta_{ij}\,, \quad
h_i = \sqrt{2\pi} i!\,, \quad
P_i(m) = m^i + {\cal O}(m^{i-1})\,,
\label{Hermite_normalize}
\end{equation}
the Vandermonde determinant can be written as
\begin{align}
\Delta(m)^2
 = \Big( \det_{ij} P_{j-1}(m_i) \Big)^2
 = 
\sum_{\sigma \tau} {\rm sgn}(\sigma \tau) 
\prod_{\ell} P_{\sigma(\ell)-1}(m_\ell) P_{\tau(\ell)-1}(m_\ell),
\end{align}
and the relevant integral is given by  
\begin{align}
& \tilde Z_k W(p)_k = 
\int \prod_n d m_n {\rm e}^{-{1\over 2}m_n^2} 
\sum_{\sigma \tau} {\rm sgn}(\sigma \tau) 
\prod_l 
P_{\sigma(l)-1}(m_l)
P_{\tau(l)-1}(m_l)
\sum_{ij} 
{1 \over N}{\rm e}^{p' m_i} 
{1 \over N}{\rm e}^{k' m_j}\,.
\label{ZW}
\end{align}
We decompose the sum over $i$, $j$ in \eqref{ZW} 
into two types according to $i=j$ or $i \neq j$ 
\begin{align}
&\eqref{ZW} \notag\\
&=
{1 \over N^2} \sum_{\sigma \tau} {\rm sgn}(\sigma \tau)
\sum_i \sum_{j \neq i} 
\int \prod_k d m_k {\rm e}^{-{1\over 2}m_k^2} 
\prod_l P_{\sigma(l)-1} (m_l) P_{\tau(l)-1} (m_l)
{\rm e}^{p' m_i + k' m_j} \label{ineqj}  \\
& \quad
+ 
{1 \over N^2} \sum_{\sigma \tau} {\rm sgn}(\sigma \tau)
\sum_i 
\int \prod_k d m_k {\rm e}^{-{1\over 2}m_k^2} 
\prod_l P_{\sigma(l)-1} (m_l) P_{\tau(l)-1} (m_l)
{\rm e}^{(p' + k') m_i}. \label{i=j}
\end{align} 
Each term \eqref{ineqj} and \eqref{i=j}
corresponds to $\tilde Z_k w(k,p)$ and 
$\tilde Z_k w(k+p)$, respectively 
where $w(k,p)$ and $w(k+p)$ are given in \eqref{W(p)kdecomp}.

The first term can be rewritten as
\begin{align}
& \eqref{ineqj} \notag \\
&=
{1 \over N^2} 
\sum_{\sigma \tau} {\rm sgn}(\sigma \tau)
\sum_i \sum_{j \neq i} \prod_{k \neq i,j} 
\int d m_k {\rm e}^{-{1 \over 2} m_k^2}
P_{\sigma(k)-1}(m_k)P_{\tau(k)-1}(m_k) \notag\\
&\qquad
\times 
\int dm_j {\rm e}^{-{1 \over 2}m_j^2}
P_{\sigma(j)-1}(m_j)P_{\tau(j)-1}(m_j) {\rm e}^{k' m_j} \notag \\
&\qquad 
\times
\int dm_i {\rm e}^{-{1 \over 2}m_i^2}
P_{\sigma(i)-1}(m_i)P_{\tau(i)-1}(m_i) {\rm e}^{p' m_i} \\
&=
{1 \over N^2} \sum_{\sigma \tau} {\rm sgn}(\sigma \tau)
\sum_i \sum_{j \neq i} 
\Big(
\prod_{k \neq i,j} \delta_{\sigma(k) \tau(k)} h_{\sigma(k)-1}
\Big)
I_{\sigma(j)-1,\tau(j)-1}(k') I_{\sigma(i)-1,\tau(i)-1}(p')\,. \label{W(p)kbyI} 
\end{align}
Here we defined 
\begin{equation}
I_{i,j} (k) \equiv
\int dm {\rm e}^{-{1 \over 2}m^2} P_i(m) P_j(m) {\rm e}^{k m}\,.
\label{I}
\end{equation}
On the other hand the second term \eqref{i=j} can be rewritten as
\begin{align}
&\eqref{i=j} \notag \\
&= 
{ 1 \over N^2 }
\sum_{\sigma \tau} {\rm sgn}(\sigma \tau)
\sum_i \int \prod_{j \neq i}
\Big(
d m_j {\rm e}^{-{1 \over 2}m_j^2}
P_{\sigma(j)-1}(m_j)
P_{\tau(j)-1}(m_j) 
\Big) \notag\\
&\quad \times 
\int d m_i {\rm e}^{-{1 \over 2}m_i^2}
P_{\sigma(i)-1}(m_i) P_{\tau(i)-1}(m_i) {\rm e}^{(p'+k') m_i} \\
&=
{1 \over N^2}
\sum_{\sigma \tau}
{\rm sgn}(\sigma \tau)
\sum_i
\Big(
\prod_{j \neq i} \delta_{\sigma(j)\tau(j)} h_{\sigma(j)-1}
\Big)
\int dm_i
{\rm e}^{-{1 \over 2} m_i^2} 
P_{\sigma(i)-1}(m_i) P_{\tau(i)-1}(m_i) {\rm e}^{(p'+k')m_i} 
\label{sgn(st)}\\
&=
{1 \over N^2}
\sum_\sigma \sum_i
\Big(
\prod_{j \neq i} h_{\sigma(j)-1}
\Big)
I_{\sigma(i)-1}(p'+k'), \label{delta(st)}
\end{align}
with $I_i(k) \equiv I_{i,i}(k)$\,. 
{}From \eqref{sgn(st)} to \eqref{delta(st)},
we used the fact that the Kronecker delta 
$\prod_{j \neq i} \delta_{\sigma(j)\tau(j)}$
implies that two permutations $\sigma$ and 
$\tau$ are identical.

Next we perform the integral in \eqref{I}.
We first note that from the generating function 
of the Hermite polynomial
\begin{equation}
{\rm e}^{t m - {t^2 \over 2}}
=
\sum_{i=0}^\infty
P_i(m) {t^i \over i!}\,,
\end{equation}
we have
\begin{equation}
P_i(m) = \partial_t^i {\rm e}^{t m - {t^2 \over 2}} \Big|_{t=0}\,.
\end{equation}
Plugging this equation into \eqref{I},
we have
\begin{align}
I_{i,j}(k)
&= 
\int dm {\rm e}^{-{1 \over 2}m^2 + km }
\partial_t^i {\rm e}^{t m - {t^2 \over 2}}
\partial_s^j {\rm e}^{s m - {s^2 \over 2}} \bigg|_{s=t=0} \notag\\
&=
\partial_t^i \partial_s^j 
{\rm e}^{-{1 \over 2}t^2} 
{\rm e}^{-{1 \over 2}s^2} 
\int dm {\rm e}^{-{1 \over 2}m^2 + (k+t+s)m}\bigg|_{s=t=0} \notag\\
&=
\sqrt{2 \pi}
\partial_t^i \partial_s^j 
{\rm e}^{-{1 \over 2}t^2} 
{\rm e}^{-{1 \over 2}s^2} 
{\rm e}^{ {1\over 2}(k+t+s)^2 } \bigg|_{s=t=0} \notag\\
&=
\sqrt{2 \pi} 
{\rm e}^{{1\over 2}k^2} \partial_t^i {\rm e}^{kt}(t+k)^j\bigg|_{t=0}.
\label{I=de}
\end{align}
By differentiating 
the generating function of Laguerre polynomial $L_i^{(\alpha)}(x)$
\begin{equation}
(1+t)^\alpha {\rm e}^{-xt} = \sum_{i=0}^\infty L_i^{(\alpha-i)}(x)t^i\,, \label{bokan}
\end{equation}
with respect to $t$ we obtain
\begin{equation}
{d^i \over d t^i}(1+t)^\alpha {\rm e}^{-xt} \bigg|_{t=0}
=
i! L_i^{(\alpha-i)}(x)\,.
\end{equation}
By using this equation with replacing 
$x \to -k^2$ and $t \to t/k$, 
we can rewrite \eqref{I=de} as
\begin{align}
I_{i,j}(k) 
=
h_i {\rm e}^{{1 \over 2}k^2} k^{j-i} L_i^{(j-i)}(-k^2)\,. \label{Iij}
\end{align}
Notice that by definition 
\begin{align}
I_{i,j}(k)=I_{j,i}(k)\,, \label{Iidentity} 
\end{align}
which can also be proved explicitly from \eqref{Iij}, 
and 
\begin{align}
I_i(k)\equiv I_{i,i}(k)=h_i{\rm e}^{{1 \over 2}k^2}L_i(-k^2), \qquad L_i(x)\equiv L_i^{(0)}(x)\,. \label{Ii}
\end{align}
Substituting \eqref{Iij} into \eqref{W(p)kbyI} yields 
\begin{align}
& \eqref{ineqj} \notag \\
&=
{1 \over N^2} \sum_{\sigma \tau} {\rm sgn}(\sigma \tau)
\sum_i \sum_{j \neq i} 
\Big(
\prod_{k \neq i,j} \delta_{\sigma(k) \tau(k)} h_{\sigma(k)-1}
\Big) \notag \\
&\qquad
\times h_{\sigma(j)-1}{\rm e}^{{1 \over 2}{k'}^2}{k'}^{\tau(j)-\sigma(j)}L_{\sigma(j)-1}^{(\tau(j)-\sigma(j))}(-{k'}^2)
           h_{\sigma(i)-1}{\rm e}^{{1 \over 2}{p'}^2}{p'}^{\tau(i)-\sigma(i)}L_{\sigma(i)-1}^{(\tau(i)-\sigma(i))}(-{p'}^2) \notag\\
&=
{1 \over N^2}{\rm e}^{{1 \over 2}({k'}^2+{p'}^2)}
\sum_{\sigma \tau} {\rm sgn}(\sigma \tau)
\Big(\prod_l h_{\sigma(l)-1}\Big)
\sum_i \sum_{j \neq i} 
\Big(
\prod_{k \neq i,j} \delta_{\sigma(k) \tau(k)}
\Big) \notag \\
&\qquad
\times {k'}^{\tau(j)-\sigma(j)}L_{\sigma(j)-1}^{(\tau(j)-\sigma(j))}(-{k'}^2)
           {p'}^{\tau(i)-\sigma(i)}L_{\sigma(i)-1}^{(\tau(i)-\sigma(i))}(-{p'}^2) \notag\\
&=
{1 \over N^2}{\rm e}^{{1 \over 2}({k'}^2+{p'}^2)}
\Big(\prod_l h_{l-1}\Big)\sum_{\sigma \tau} {\rm sgn}(\sigma \tau)
\sum_i \sum_{j \neq i} 
\Big(
\prod_{k \neq i,j} \delta_{\sigma(k) \tau(k)}
\Big) \notag \\
&\qquad
\times {k'}^{\tau(j)-\sigma(j)}L_{\sigma(j)-1}^{(\tau(j)-\sigma(j))}(-{k'}^2)
           {p'}^{\tau(i)-\sigma(i)}L_{\sigma(i)-1}^{(\tau(i)-\sigma(i))}(-{p'}^2)\,.            
\end{align}
Here the Kronecker delta in the last expression implies that 
only two cases are possible: 
\begin{enumerate}
\item $\sigma=\tau$,
\item $\sigma(k)=\tau(k)$ for $k\neq i,j$ 
and $\sigma(i)=\tau(j)$, $\sigma(j)=\tau(i)$. 
\end{enumerate}
According to this, we get
\begin{align}
& \eqref{ineqj} \notag \\
&=
{1 \over N^2}{\rm e}^{{1 \over 2}({k'}^2+{p'}^2)}
\Big(\prod_l h_{l-1}\Big)\sum_{\sigma} 
\sum_i \sum_{j \neq i} \notag \\
&\qquad
\times \Big(L_{\sigma(j)-1}(-{k'}^2)L_{\sigma(i)-1}(-{p'}^2)
               -{k'}^{\sigma(i)-\sigma(j)}L_{\sigma(j)-1}^{(\sigma(i)-\sigma(j))}(-{k'}^2)
                 {p'}^{\sigma(j)-\sigma(i)}L_{\sigma(i)-1}^{(\sigma(j)-\sigma(i))}(-{p'}^2)\Big) \notag\\
&=
{N! \over N^2}{\rm e}^{{1 \over 2}({k'}^2+{p'}^2)}\Big(\prod_l h_{l-1}\Big) \notag\\
&\qquad
\times \sum_{ij} 
           \Big(L_{j-1}(-{k'}^2)L_{i-1}(-{p'}^2)
               -{k'}^{i-j}L_{j-1}^{(i-j)}(-{k'}^2)
                 {p'}^{j-i}L_{i-1}^{(j-i)}(-{p'}^2)\Big) \notag\\
&=
{N! \over N^2}{\rm e}^{{1 \over 2}({k'}^2+{p'}^2)}\Big(\prod_l h_{l-1}\Big) \notag\\
&\qquad
\times \Big(L_{N-1}^{(1)}(-{k'}^2)L_{N-1}^{(1)}(-{p'}^2)
               -\sum_{ij}{k'}^{i-j}L_{j-1}^{(i-j)}(-{k'}^2)
                            {p'}^{j-i}L_{i-1}^{(j-i)}(-{p'}^2)\Big)\,, \label{w(k,p)}                
\end{align}
where in the last step we have used an identity of the Laguerre
polynomial,
$L_n^{(\alpha+1)}(x)=\sum_{j=0}^n L_j^{(\alpha)}(x)$.

On the other hand, using \eqref{Ii} in \eqref{delta(st)}, we obtain 
\begin{align} 
&\eqref{i=j} \notag \\
&={1 \over N^2}
\sum_\sigma \sum_i
\Big(
\prod_{j \neq i} h_{\sigma(j)-1}
\Big)
h_{\sigma(i)-1}{\rm e}^{{1 \over 2}(p'+k')^2}L_{\sigma(i)-1}(-(p'+k')^2) \notag \\
&={1 \over N^2}{\rm e}^{{1 \over 2}(p'+k')^2}
\sum_\sigma \sum_i
\Big(
\prod_j h_{\sigma(j)-1}
\Big)
L_{\sigma(i)-1}(-(p'+k')^2) \notag \\
&={1 \over N^2}{\rm e}^{{1 \over 2}(p'+k')^2}
\Big(
\prod_j h_{j-1}
\Big)
\sum_\sigma \sum_i
L_{\sigma(i)-1}(-(p'+k')^2) \notag \\
&={N! \over N^2}{\rm e}^{{1 \over 2}(p'+k')^2}
\Big(
\prod_j h_{j-1}
\Big)
\sum_i L_{i-1}(-(p'+k')^2) \notag \\
&={N! \over N^2}{\rm e}^{{1 \over 2}(p'+k')^2}
\Big(
\prod_j h_{j-1}
\Big)
L_{N-1}^{(1)}(-(p'+k')^2)\,. \label{w(p+k)}
\end{align}
{}From \eqref{w(k,p)} and \eqref{w(p+k)}, \eqref{ZW} can be rewritten as 
\begin{align}
\tilde Z_k W(p)_k 
=&
{N! \over N^2}{\rm e}^{{1 \over 2}({k'}^2+{p'}^2)}\Big(\prod_l h_{l-1}\Big) \notag\\
\times & \Big(L_{N-1}^{(1)}(-{k'}^2)L_{N-1}^{(1)}(-{p'}^2)
               -\sum_{ij}{k'}^{i-j}L_{j-1}^{(i-j)}(-{k'}^2)
                            {p'}^{j-i}L_{i-1}^{(j-i)}(-{p'}^2) \notag\\
&\quad
+{\rm e}^{k'p'}L_{N-1}^{(1)}(-(k'+p')^2)\Big)\,.                            
\end{align}
Moreover, by setting $p=0$ in \eqref{ZW} and noting $W(0)_k=1$,  
it is easy to see that $\tilde Z_k$ is nothing but \eqref{i=j} with $p'=0$ 
multiplied by $N$. Thus from \eqref{w(p+k)}
\begin{align}
\tilde Z_k={N! \over N}{\rm e}^{{1 \over 2}{k'}^2}
\Big(
\prod_j h_{j-1}
\Big)
L_{N-1}^{(1)}(-{k'}^2)\,.
\end{align}
Therefore we find that 
\begin{align}
W(p)_k 
=&
{1 \over N}{\rm e}^{{1 \over 2}{p'}^2}L_{N-1}^{(1)}(-{p'}^2)
-{1 \over N}{\rm e}^{{1 \over 2}{p'}^2}\sum_{ij}\frac{{k'}^{i-j}L_{j-1}^{(i-j)}(-{k'}^2)
                            {p'}^{j-i}L_{i-1}^{(j-i)}(-{p'}^2)}{L_{N-1}^{(1)}(-{k'}^2)} \label{w(k,p)final} \\
+&
{1 \over N}\frac{{\rm e}^{{1 \over 2}(k'+p')^2}L_{N-1}^{(1)}(-(k'+p')^2)}{{\rm e}^{{1 \over 2}{k'}^2}L_{N-1}^{(1)}(-{k'}^2)}                            
\,. \label{w(k+p)final}
\end{align}
\eqref{w(k,p)final} and \eqref{w(k+p)final} come from \eqref{ineqj} and \eqref{i=j}, 
so give the explicit form of $w(k,p)$ and $w(k+p)$ in \eqref{W(p)kdecomp}, respectively. 
Notice that so far we have not made any approximation and hence this equation 
is exact and holds for any finite $N$. Therefore we can expect its useful application elsewhere. 

\subsection{The evaluation of the second term in $w(k,p)$}
\label{sec:eval-secondterm}

In this subsection, we evaluate the large-$N$ limit 
of the second term in \eqref{w(k,p)final} 
and derive the corresponding term in the resolvent. 
We consider the numerator of the second term:
\begin{align}
A &= \sum_{i,j=0}^{N-1} F_{i,j}(k',p'), \label{A=sumF} \\
F_{i,j} (k',p') &=
{\rm e}^{k'^2 \over 2} k'^{i-j} L_j^{(i-j)}(-k'^2)
{\rm e}^{p'^2 \over 2} p'^{j-i} L_i^{(j-i)}(-p'^2) \\
& 
= 
h_j^{-1} I_{j,i}(k')
h_i^{-1} I_{i,j}(p'),
\label{F=II}
\end{align}
where we have multiplied 
the additional factor ${\rm e}^{k'^2 \over 2}$ and, 
for notational simplicity, we have changed 
$i \to i+1$, $j \to j+1$.
$h_i$ and $I_{i,j}$ are defined as 
\eqref{Hermite_normalize}
and 
\eqref{I},
respectively.

Let us start with the large-$N$ expansion of 
$I_{j,i}(k')$ for arbitrary $i$, $j$ and 
finite $\kappa$. 
For this purpose, we recall its integral representation 
\eqref{I} and rescale the variable as 
$ \tilde m = m/\sqrt{N} $ such that 
the Gaussian potential becomes proportional to $N$;
\begin{equation}
I_{j,i}(k') 
= 
N^{i+j+1 \over 2}
\int d \tilde m
{\rm e}^{- {N \over 2} \tilde m^2 + \tilde k' \tilde m}
\widetilde P_j(\tilde m) \widetilde P_i(\tilde m).
\label{I_rescale}
\end{equation}
Here $\tilde k' = \sqrt{N} k'$ and 
we have also rescaled the Hermite polynomials 
as $\widetilde P_i(\tilde m) = N^{-i/2} P_i(m)$
to make it satisfy the normalization 
$\widetilde P_i(\tilde m) = \tilde m^i + \cdots$.
The large-$N$ behavior of the 
generic orthogonal polynomial in this normalization
is addressed in \cite{Hanada:2004im}.
Because of the large linear term 
$\tilde k' \tilde m = 2 \kappa N \tilde m$, 
integral \eqref{I_rescale} has a saddle point $\tilde m = \tilde m_\ast$
at a non-oscillating region $\sqrt{4i/N} < \tilde m_\ast$ 
for the Hermite polynomial $\widetilde P_i(\tilde m)$ 
(see \cite{Hanada:2004im} for generic behavior of the orthogonal polynomial).  
Hence we concentrate on the expression for that region:
\begin{equation}
\widetilde P_i( \tilde m)
=
\exp
\bigg(
N \int_0^{i \over N} d \xi \log k^{(0)} (\tilde m ,\xi)
+
{1 \over 2} \log k^{(0)}\Big( \tilde m, {i \over N} \Big)
-
{1 \over 2} \log q \Big( \tilde m, { i \over N} \Big)
+
{\cal O}(N^{-1})
\bigg), \label{asymP}
\end{equation}
where 
\begin{equation}
k^{(0)}(\tilde m, \xi) 
= 
{ \tilde m + \sqrt{\tilde m^2 - 4 \xi} \over 2}, \quad
q(\tilde m, \xi)
=
\sqrt{\tilde m^2 - 4 \xi}\,.
\end{equation}
By using this expression, we find the saddle point 
of the integral \eqref{I_rescale} as 
\begin{equation}
\tilde m_\ast^2 
= 
{\tilde k'^2 \over N^2}
+
{2 (i+j) \over N}
+
{(i-j)^2 \over \tilde k'^2}.
\end{equation}
Evaluating \eqref{I_rescale} semi-classically 
including the Gaussian integral around the saddle point,
we obtain the following large-$N$ behavior
\begin{align}
I_{j,i}(k')
&=
N^{i+j+1 \over 2}
\exp
\Bigg[
{1 \over 2} \tilde k' \tilde m_\ast
+ i \log 
{ 
\tilde m_\ast + {\tilde k' \over N} + {j-i \over \tilde k'}
\over 2
}
+ j \log 
{ 
\tilde m_\ast + {\tilde k' \over N} + {i-j \over \tilde k'}
\over 2
}
-{i \over 2} - {j \over 2} \notag\\
&\hspace{1cm}+
{1 \over 2} 
\log
{
\Big( \tilde m_\ast + {\tilde k' \over N} \Big)^2 
- 
{(i-j)^2 \over \tilde k'^2}
\over 
4
}
-
{1 \over 2} \log (\tilde k' \tilde m_\ast)
+
{1 \over 2} \log (2 \pi) + {\cal O}(N^{-1})
\Bigg].
\label{I(k)}
\end{align}

Next we study the large-$N$ behavior of $I_{i,j}(p')$ 
for finite $p = p' \sqrt{4N / \lambda} $. 
In this case, the linear term $\tilde p' \tilde m = p' m$ 
is small and we need to consider an oscillating region.
Here, instead of studying it, 
we derive the large-$N$ behavior of $I_{i,j}(p')$
by solving the differential equation satisfied by
$f(i,j,p') = {\rm e}^{p'^2 \over 2} L_i^{(j-i)}(-p'^2)$:
\begin{equation}
\partial_p^2 f
+
{2 ( j - i ) + 1 \over p}
\partial_p f
-
{\lambda \over 4N}
\bigg(
{\lambda \over 4N} p^2
+
2 ( i + j + 1)
\bigg)
=
0.
\end{equation}
The first term $(\lambda/4N)p^2 $ 
in the round bracket is small compared 
to the second term in the same bracket.
By neglecting this term, 
we can reduce the the equation 
to Bessel's differential equation 
and we obtain 
\begin{equation}
I_{i,j}(p') 
= 
\sqrt{2 \pi } {\rm max}(i,j)!
\bigg(
{2 \over i + j + 1}
\bigg)^{|i-j| \over 2}
I_{|i-j|} (p \sqrt{\Lambda}) 
\Big(
1 + {\cal O}(N^{-1})
\Big),
\label{I(p)}
\end{equation}
where $\Lambda = {\lambda ( i + j+ 1) / 2N}$.
 ${\rm max}(i,j)$ represents the larger value
and $I_{|i-j|}$ is the modified Bessel function.
The overall constant can be fixed by comparing the
asymptotic behavior of the Bessel function with that of the 
Laguerre polynomial in the small $p$ limit.
This expression is again valid for any values of $i$ and $j$.

Now we discuss the summation \eqref{A=sumF}
using these ingredients.
We should be careful that the subleading ${\cal O}(N^{-1})$
terms in \eqref{I(k)} and \eqref{I(p)}
possibly contribute to the leading behavior of $A$, 
since $A$ is defined by the summation 
over $N^2$ terms.
However, as we will see later,
only the terms with $N-i={\cal O}(1)$ 
and $N-j={\cal O}(1)$ actually contribute 
to the second term of \eqref{w(k,p)final}.
Then the effective number of terms being summed is of order 
${\cal O}(1)$,
and we can consistently neglect the ${\cal O}(N^{-1})$ terms.
In the following, we call the range $N-i={\cal O}(1)$,
$N-j={\cal O}(1)$ ``${\cal C}$''.
We first concentrate on the terms in the range ${\cal C}$, 
and then we study the contributions of the terms outside ${\cal C}$.

In order to discuss the terms in ${\cal C}$,
we change variables as
$ m \equiv 2N - (i + j)$ and $d \equiv i-j$ 
and expand \eqref{I(k)} and \eqref{I(p)}
with assuming $m={\cal O}(1)$, $d = {\cal O}(1)$.
Then we obtain the following simple expression for 
$F_{i,j}(k',p')$:
\begin{align}
& F_{i,j}(k',p') 
=
\Bigg[
2N \Big( \kappa \sqrt{\kappa^2+1} 
+ \log(\sqrt{\kappa^2 + 1} + \kappa) \Big)
+ \log(\sqrt{\kappa^2 + 1} + \kappa ) \\
&\hspace{2cm} - {1 \over 2} \log(\kappa \sqrt{\kappa^2 + 1} )
- {1 \over 2} \log (8\pi N)
\Bigg]
\,\,{
I_{|d|}( p \sqrt{\lambda} )
\over 
\big( \sqrt{\kappa^2 + 1} + \kappa \big)^{m} 
}\,\,
\Big(1 + {\cal O}(N^{-1})\Big).
\label{F(m,d)}
\end{align}

We set the range of summation as 
$d = 0, \pm 1, \pm 2 , \cdots, \pm D$
and $m = |d| + 2, |d| + 4, \cdots, |d| + 2 M$
for a fixed value of $d$.
Here $M$ and $D$ define the upper limits 
of the summation which should be large 
but still of  ${\cal O}(N^0)$ in the present 
assumption of the range ${\cal C}$.
In the large-$N$ limit, both of the
upper limits can be consistently taken to be infinity. 
After the $m$-summation, we obtain the following 
large-$N$ expression for $A$:
\begin{align}
A 
&= 
\sum_{d=-D}^D
\sum_{(m-|d|)/2 = 1}^M
F_{i,j}(k',p') 
+
(\cdots)
\label{Asum1}\\
& \to
{\rm e}^{k'^2 \over 2}
L_{N-1}^{(1)}(-k'^2) 
\sum_{d=-\infty}^\infty
{I_{|d|}(p\sqrt{\lambda}) \over (\sqrt{\kappa^2 + 1} + \kappa)^{|d|}}
\Big( 1 + {\cal O}(N^{-1}) \Big)
+ (\cdots).
\label{Asum2}
\end{align}
Here for simplicity we have taken the limit 
$D \to \infty$, $M \to \infty$ 
which is consistent in the large-$N$ limit
as noted above.
The extracted factor 
${\rm e}^{k'^2 \over 2} L_{N-1}^{(1)}(-k'^2)$ 
is exactly canceled by the denominator of 
\eqref{w(k,p)final} taking account of the extra factor
${\rm e}^{k'^2 \over 2}$.

The dots ($\cdots$) in \eqref{Asum2} express the 
terms outside ${\cal C}$.
We will shortly see that 
these terms actually dose not 
give finite contribution to the resolvent
in the large-$N$ limit.
Before that 
we shall derive the resolvent corresponding to 
the first term 
of \eqref{Asum2} by performing the 
Laplace transformation (recall the overall factor $-1/N$ 
in the second term of \eqref{w(k,p)final}):
\begin{align}
R_\textrm{2nd term}(z)
&= -{1 \over N}
\int_0^\infty dp {\rm e}^{-p z}
\sum_{d=-\infty}^{\infty}
{
I_{|d|} (p \sqrt{\lambda})
\over 
(\sqrt{\kappa^2 + 1} + \kappa)^{|d|}
} \\
& =
-{1 \over N}
{1 \over \sqrt{z^2 - \lambda}}
\Bigg(
1 + 2 \sum_{d=1}^\infty
\Bigg(
{z - \sqrt{z^2 - \lambda} 
\over 
\sqrt{\lambda}(\kappa + \sqrt{\kappa^2 + 1})}
\Bigg)^d
\Bigg) \\
&=
-{1 \over N}
{z + \sqrt{\lambda}\sqrt{1 + \kappa^2}
\over
\sqrt{z^2 - \lambda} (\sqrt{z^2-\lambda}+\sqrt{\lambda}\kappa) }.
\label{R2nd}
\end{align}
Therefore, the only singularity of $R_\textrm{2nd term}(z)$ 
is the cut $-\sqrt{\lambda} \leq z \leq \sqrt{\lambda}$.
In fact, it is easy to see that \eqref{R2nd}
is just another form of \eqref{R(0,1)}.
By checking the residue of the pole at infinity
it is clear that the resolvent 
corresponds to ``minus one'' eigenvalue, i.e.,
it subtracts one eigenvalue from the leading semi-circle distribution.
Hence we have derived the resolvent \eqref{R(0,1)}
without assuming the conditions mentioned after \eqref{R(0,1)def}.

Finally let us show 
that the terms outside ${\cal C}$
do not contribute to the resolvent.
In fact these terms are exponentially suppressed 
with respect to $N$ compared to the terms in the range 
${\cal C}$. 
Since such exponential suppressions will not 
be compensated even by summing up all the $N^2$ terms,
we can conclude that these terms do not contribute
in the large-$N$ limit. 

In order to clarify this point, 
it is helpful to start with 
the following simple 
ordering property of $F_{i,j}(k',p')$:
\begin{equation}
F_{i,j}(k',p') \leq F_{i+1,j+1}(k',p'), \label{order}
\end{equation}
which follows directly from the definition 
of the Laguerre polynomial: 
$L_i^{(j-i)}(-x^2) = \sum_{r=0}^i {}_j C_{i-r} x^{2r}/r!$.
Here ${}_j C_{i-r}$ is a binomial coefficient and 
we define ${}_j C_{i-r} = 0$ for $j-i+r < 0$.
Because of this ordering property and also the symmetric 
property $F_{i,j}(k',p') = F_{j,i}(k',p')$, 
it is sufficient to show the following two claims:
(I) among the terms $F_{i,N-1}(k',p')$, 
those in ${\cal C}$ are exponentially larger 
than the other terms, and 
(II) among the terms $F_{i,j}(k',p')$ with 
$i-j={\cal O}(1)$, those in ${\cal C}$
are exponentially larger than the other terms. 
By showing these two claims and considering the 
ordering property \eqref{order}, 
one can easily see that the terms in ${\cal C}$ are 
exponentially larger than all the other terms.

Let us begin with the first claim (I).
We use the expressions \eqref{I(k)} 
and \eqref{I(p)} with setting $j=N-1$.
Keeping terms of up to ${\cal O}(N)$,
we have
\begin{align}
& F_{i,N-1}(k',p')
=
\exp\Bigg[
{N \over 2}\big(\xi-1 \big) \log N 
+
{N \over 2}\big( 1 -\xi )
+
{N \over 2}  R 
+
N \log {K_- \over 2}
+
N \xi \log {K_+ \over 2} 
\notag
\\
& \hspace{2.5cm}
+ N\big(1- \xi \big) \log p'
-N \xi \log \xi
- N\big( 1-\xi \big) \log \big( 1 - \xi \big)
+ {\cal O}(\log N)
\Bigg]\,,\label{F(i,N-1)}
\end{align}
where $\xi = i/N$ and $K_\pm$ and $R$ are defined as
\begin{equation}
K_\pm = 
{R \over 2 \kappa} 
+ 2 \kappa 
\pm 
{1 \over 2 \kappa} (1 - \xi)\,,\quad 
R = \sqrt{16 \kappa^4 + 8 \kappa^2 (1+\xi) + (1-\xi)^2}.
\end{equation}
The first and the second line is the asymptotic form 
of $h_{N-1}^{-1} I_{N-1,i}(k')$ and $h_i^{-1} I_{i,N-1}(p')$,
respectively. We have used the asymptotic form of the 
Bessel function which follows from the saddle point 
approximation of the integral:
\begin{align}
h_i^{-1} I_{i,N-1}(p')
&=p'^{N-1-i} {(N-1)! \over i!} 
\bigg( {2 \over p \sqrt{\Lambda}} \bigg)^{N-1-i}
I_{N-1-i}(p \sqrt{\Lambda})\Big( 1 + {\cal O}(N^{-1}) \Big)\\
&=
{(N-1)! \over i!} {p'^\nu \over \sqrt{\pi} \Gamma(\nu + 1/2)}
\int_{-1}^1 dt
(1 - t^2)^{\nu - {1 \over 2}} {\rm e}^{-p \sqrt{\Lambda}t}
\Big( 1 + {\cal O}(N^{-1}) \Big),
\label{Bessel=int}
\end{align}
where $\nu = N - 1 - i $.
The second line of \eqref{F(i,N-1)}
follows by assuming $\nu \gg 1 $, 
and applying the saddle point approximation 
for the integral 
and also Stirling's formula for $\Gamma(\nu + 1/2)$.
For the range $\nu = {\cal O}(1)$, 
in the sense of large-$N$ limit, 
both of these approximations break down. 
However, even in this region, 
we can find the regime where 
$\nu$ is still large compared 
to $1$ and also $p \sqrt{\lambda}$.
In such a regime, the asymptotic 
behavior \eqref{F(i,N-1)} is still valid.
In the second line of \eqref{F(i,N-1)}, 
we have also used Stirling's formula for $i!$.
It is easy to check that the following
arguments are still intact even if we 
consider the regime $i={\cal O}(1)$
seriously.

The dominant $\xi$-dependence of 
\eqref{F(i,N-1)} arises from 
the first term on the first line 
and the terms on the second line.
Let us take the derivative 
of the summation of these dominant terms
with respect to $\xi$:
\begin{equation}
{\partial \over \partial \xi}
N \bigg(
{1 \over 2}\xi \log N
-
 \xi \log p'
- \xi \log \xi
-
(1-\xi) \log ( 1 - \xi )
\bigg)
=
N \log {N^\gamma \over 1 - N^{\gamma-1}} 
+
{\cal O}(N)\,.
\end{equation}
In the right hand side, we introduced $\gamma$
by $N-i=N^\gamma$, i.e., $1 - \xi = N^{\gamma-1}$ 
and used $p'\simeq N^{-1/2}$.
From this expression, we see that 
for the range $ 0 < \gamma \leq 1$,
$F_{i,N-1}(k',p')$ increases quite rapidly 
with respect to $\xi$ as $N^{\gamma N \xi}$.

Let us also check the $\xi$-dependence 
for the range $1-\xi={\cal O}(N^{-1})$, i.e.,
$N - i = {\cal O}(1)$.
For this purpose we expand \eqref{F(i,N-1)}
with respect to the small parameter 
$ \epsilon = 1 - \xi $.
Then we have
\begin{equation}
F_{i,N-1}(k',p')
=
\exp N
\bigg(
-
\epsilon \log (\kappa + \sqrt{1+\kappa^2})
-
{1 \over 2} \epsilon \log N
+
\epsilon \log p' + \epsilon - \epsilon \log \epsilon
+ \cdots
\bigg).
\end{equation}
Here we have written down only the leading 
$\epsilon$-dependence.
So, dots ($\cdots$) include the subleading terms 
and also the leading but $\epsilon$-independent terms. 
By taking the derivative of the exponent 
with respect to $\xi = 1- \epsilon$, 
we obtain the following result:
\begin{equation}
N \log (\kappa + \sqrt{\kappa^2 + 1 }) 
+ 
N \log \bigg( {2 \over p \sqrt{\lambda}} N \epsilon \bigg).
\end{equation}
So, for the range $\epsilon > p \sqrt{\lambda} / 2 N$,
the function $F_{i,N-1}(k',p')$ increases with $\xi$ 
at least as 
${\rm e}^{\xi N \log (\kappa + \sqrt{\kappa^2 + 1})}$.
The ${\cal O}(\log N)$ terms in \eqref{F(i,N-1)} 
do not change this conclusion.
Hence we have shown the first claim (I).

Next we turn to the second claim (II), i.e.,
the parameter range $i-j = {\cal O}(1)$.
For this range, $h_i^{-1} I_{i,j}(p')$
behaves, at most, like power in $N$ as can be seen from \eqref{I(p)}.
Hence the dominant behavior is determined by
the following asymptotic form of \eqref{I(k)}:
\begin{equation}
{\rm e}^{k'^2 \over 2} k'^{i-j} L_j^{(i-j)}(-k'^2) 
= 
\exp
\Bigg( 
2N 
\bigg( 
\kappa \sqrt{\kappa^2 + \zeta} 
+ 
\zeta \log 
{
\kappa + \sqrt{\kappa^2 + \zeta}
\over \sqrt{ \zeta }
}
\bigg)
+
{\cal O}(\log N)
\Bigg),
\end{equation}
where $\zeta = j / N$.
The dependence on $i-j$ appears only 
in the subleading terms.
By taking derivative of the leading exponent 
with respect to $\zeta$,
we find that $F_{i,j}(k',p')$ in this range
grows at least as 
$F_{i,j}(k',p')
\sim 
{\rm e}^{2N\zeta \log (\kappa + \sqrt{\kappa^2 + 1})}$.
This shows that the second claim (II) indeed holds.

\section{BPS conditions and the uniqueness of the solution}
\label{App:BPS-cond}

In AdS/CFT correspondence, one of the most important guiding principles to
find out a corresponding object to its holographic counterpart is the
symmetry preserved by the object.
For example, the circular and the straight line Wilson loops
preserve $SL(2,\bR)\times SO(3)$ symmetry as a part of the Euclidean
conformal group $SO(5,1)$ of the ${\cal N}=4$ super Yang-Mills (SYM) theory.
In accordance with this preserved symmetry, the D3-brane solutions
found in \cite{Drukker:2005kx} have the structure of AdS$_2 \times$
S$^2$ whose isometry is indeed  $SL(2,\bR)\times SO(3)$\footnote{%
The small fluctuation around the loop also obeys the classification
with respect to this symmetry. See, for example, \cite{DK}.}.
The circular loop and the straight line also preserve a part of
the supersymmetry of SYM, namely they are BPS objects, and therefore
corresponding D3-brane
solutions should preserve some of global supersymmetry of type IIB
supergravity and, as also shown in \cite{Drukker:2005kx},
so do they.

In this appendix, we see that the BPS condition for D3-brane solutions,
together with the S$^2$ symmetric ansatz, suffices to determine the
solution (virtually) uniquely, at least in the circular loop and the
straight line cases.
In \cite{Drukker:2005kx}, the authors checked that, for the straight
line, the BPS equation is satisfied by their solution.
Here we do not assume the D3-brane solution, but just postulate the S$^2$
symmetric ansatz $\eta=\eta(\rho)$ and $F_{\rho\psi}(\rho)$, and
observe how the BPS condition restricts the form of the solution.

We examine the circular loop case with a brief summary of the
BPS condition for D-brane solutions.
We start with the Euclidean AdS$_5$ metric (\ref{circlemetric}),
\begin{align}
\label{eq:circle-metric}
  ds^2=&
\frac{L^2}{\sin^2 \eta } \left(
d\eta^2
+\cos^2 \eta \, d\psi^2
+d\rho^2
+\sinh^2 \rho \, (d\theta^2 + \sin^2 \theta \, d\phi^2 )
\right) \,,
\end{align}
and the S$^5$ part is omitted in the analysis since the solutions we
are interested in are trivial on S$^5$.
By a coordinate transformation, this metric is mapped into 
\begin{equation}
  ds^2= \frac{L^2}{y^2}\left(
dy^2
+dt^2
+dr^2
+r^2(d\theta^2 + \sin^2 \theta d\phi^2)
\right) \,,
\end{equation}
where Wick rotation to Lorentzian metric, $t_E=i t$, is easily
understood.
One can see that this Wick rotated Lorentzian metric is mapped into
the metric (\ref{eq:circle-metric}) with $\psi$ replaced with
$\psi_E=i\psi$,
and then we can consider $\psi_E=i\psi$ as ``Wick rotation'' of our
metric.
From (\ref{eq:circle-metric}), we define the vielbeins,
\begin{align}
  e^1 =& \frac{L}{\sin \eta} d\eta \,, \quad
  e^2 = \frac{L}{\sin \eta} \cos\eta d\psi \,, \quad
  e^3 = \frac{L}{\sin \eta} d\rho \,,
\nn\\
  e^4 =& \frac{L}{\sin \eta} \sinh\rho d\theta \,, \quad
  e^5 = \frac{L}{\sin \eta} \sinh\rho \sin \theta d\phi \,.
\end{align}
We then consider the Killing spinor equation,
\begin{align}
  \left(D_\mu +\frac{1}{2L}\Gamma_\star \gamma_\mu \right) \epsilon=0 \,,
\end{align}
where $\Gamma_\star = \Gamma_{12345}$ and the gamma matrices
in the flat tangent space are denoted by capital $\Gamma_a$ and those in the
curved background by $\gamma_\mu = e^a_\mu \Gamma_a$, and
$\epsilon$ is a complex combination of two chiral Majorana-Weyl
spinors of type IIB supergravity.
We find the Killing spinor of this background,
\begin{align}
\label{KillingSpinor-Circle}
  \epsilon=&
\tan^{1/2} \left(\frac{\eta}{2} \right)
 (M_+ \epsilon_1^- + M_- \epsilon_2^-)
+ i \cot^{1/2} \left(\frac{\eta}{2} \right)
 \Gamma_{12} (M_+ \epsilon_1^- - M_- \epsilon_2^-)\,,
\end{align}
where $\epsilon_{1,2}^-$ are constant spinors satisfying $\tilde\Gamma 
\epsilon_{1,2}^- = - \epsilon_{1,2}^-$,
$\tilde\Gamma=\Gamma_{2345}$, and
\begin{equation}
  M_\pm (\psi,\rho,\theta,\phi)=
e^{\pm \frac{i}{2}\psi} 
e^{\pm \frac{i\rho}{2} \Gamma_{23}} e^{\frac{\theta}{2}\Gamma_{34}}
e^{\frac{\phi}{2}\Gamma_{45}} \,.
\end{equation}

The BPS condition for a D-brane solution is given by a compatibility
condition of the $\kappa$-symmetry of the D-brane action and a part of the
global supersymmetry of the background.
See, for example, \cite{Skenderis:2002vf}.
The $\kappa$-symmetry projector with Lorentzian signature is defined by
\begin{align}
  d^{p+1}\xi \Gamma =&
 -e^{-\phi} \mathcal{L}_{DBI}^{-1} \, e^{\mathcal{F}} \wedge X
 \big|_\text{vol}  \,,
\\
X \equiv & \oplus \Gamma_{(2n)} K^n I \,,
\\
\Gamma_{(n)}=& \frac{1}{n!}d\xi^{i_1} \wedge \cdots \wedge d\xi^{i_n}
 \partial_{i_1} X^{\mu_1}
\cdots \partial_{i_n} X^{\mu_n} \gamma_{\mu_1 \cdots \mu_n} \,,
\end{align}
where $\xi$ denotes the world volume coordinates, $X^\mu(\xi)$ are
the embedding coordinates, $\mathcal{F}=2\pi\alpha' F$ is the two-form
field strength of the world volume $U(1)$ gauge field,
and $K$ and $I$ act on spinors as $K \psi= \psi^*, I\psi=-i
\psi$.
This projector is traceless and equipotent,
\begin{equation}
  \Tr \Gamma=0 \,, 
\qquad
\Gamma^2=1 \,,
\end{equation}
and the D-brane action enjoys the $\kappa$-symmetry
\begin{equation}
  \delta \theta(\xi)= (1+\Gamma (\xi) ) \kappa(\xi) \,,
\end{equation}
where $\theta$ is the fermionic partner of $X^\mu$.
$\theta$ is fermionic coordinate of the target space-time as well
and then it transforms under the space-time supersymmetry as
$\delta\theta=\epsilon$.
Together with the $\kappa$-symmetry, $\theta$ transforms as
\begin{equation}
\delta  \theta= (1+\Gamma) \kappa + \epsilon \,,
\end{equation}
and therefore if a constant $\epsilon$ satisfies
\begin{equation}
\label{BPS-eq-Dbrane}
  (1-\Gamma) \epsilon =0 \,,
\end{equation}
then this global space-time supersymmetry is compatible with the
$\kappa$-symmetry, hence the embedded D-brane is BPS.

We choose $\rho,\psi,\theta,\phi$ as the world volume coordinates as before,
and then with the S$^2$ symmetric ansatz, the projector takes the
following form
after Wick rotation,
\begin{align}
  \Gamma
=&
 i f(\eta,F)^{-1} \left[
\cos\eta (1+\eta' \Gamma_{13})
- 2\pi \alpha' \frac{\sin^2 \eta}{L^2} F_{\psi\rho} \Gamma_{23} K
\right] \bar\Gamma I \,,
\end{align}
where
\begin{align}
  f=&\sqrt{\cos^2 \eta (1+ \eta^{\prime 2}) + (2\pi \alpha')^2 \frac{\sin^4
  \eta }{L^4} F^2_{\psi\rho}}
 \,.
\end{align}
Now we try to solve the BPS equation
\begin{equation}
  \Gamma\epsilon=\epsilon \,.
\end{equation}
First we rewrite the Killing spinor
(\ref{KillingSpinor-Circle}),
\begin{align}
  \epsilon=&
T(\eta) (M_+ \epsilon_1^- + M_- \epsilon_2^-)
\nn\\&
+ i \cosh \rho C(\eta) (M_+ \Gamma_{12} \epsilon_1^- - M_-
\Gamma_{12} \epsilon_2^-)
\nn\\&
+ \sinh\rho C(\eta) (M_+ g(\theta,\phi) \Gamma_{1} \epsilon_1^-
  + M_- g(\theta,\phi) \Gamma_{1} \epsilon_2^-) \,,
\end{align}
where
\begin{align}
g(\theta,\phi) \equiv &
\Gamma_3 \cos \theta +\Gamma_4 \sin\theta \cos\phi 
+ \Gamma_5 \sin\theta \sin\phi \,,
\end{align}
and $C(\eta) \equiv \cot^{1/2}
\left(\frac{\eta}{2} \right), T(\eta) \equiv \tan^{1/2}
\left(\frac{\eta}{2} \right)$ have been introduced for simplicity.
The action of the projector $\Gamma$ on the Killing spinor is
\begin{align}
 i f(\eta,F) \Gamma \epsilon =&
-i \cos\eta T(\eta) (M_+ \epsilon_1^- + M_- \epsilon_2^-)
\nn\\&
-\cos\eta \left( \cosh \rho C(\eta) + \eta' \sinh\rho T(\eta) \right)
 (M_+ \Gamma_{12} \epsilon_1^- - M_- \Gamma_{12} \epsilon_2^-) 
\nn\\&
+ i\cos\eta \left( \sinh\rho C(\eta) + \eta'\cosh\rho T(\eta) \right)
 (M_+ g(\theta,\phi) \Gamma_{1} \epsilon_1^-  + M_- g(\theta,\phi) \Gamma_{1} \epsilon_2^-)
\nn\\&
+\eta' \cos\eta C(\eta) 
 (M_+ g(\theta,\phi) \Gamma_{2} \epsilon_1^-  - M_- g(\theta,\phi) \Gamma_{2} \epsilon_2^-)
\nn\\&
+i 2\pi \alpha' \frac{\sin^2 \eta}{L^2} F_{\psi\rho} \left(
- T(\eta)
 ( M_+ g(\theta,\phi) \Gamma_2 \bar\Gamma \epsilon_1^{-*} +M_-
 g(\theta,\phi) \Gamma_2 \bar\Gamma \epsilon_2^{-*})
\right.\nn\\& \left. \hskip1.8cm
- i C(\eta) \cosh\rho 
 ( M_+ g(\theta,\phi) \Gamma_1 \bar\Gamma \epsilon_1^{-*} - M_-
 g(\theta,\phi) \Gamma_1 \bar\Gamma \epsilon_2^{-*})
\right.\nn\\& \left. \hskip1.8cm
+ C(\eta) \sinh\rho 
 ( M_+ \Gamma_1 \Gamma_2 \bar\Gamma \epsilon_1^{-*} + M_- \Gamma_1
 \Gamma_2 \bar\Gamma \epsilon_2^{-*}) 
\right) \,.
\end{align}
In order for a solution to exist, $\epsilon_{1,2}^-$ have to be
related to their complex conjugates $\epsilon_{1,2}^{- *}$ in a certain
way, say,
\begin{equation}
\Gamma_{1,2}  \Bar\Gamma \epsilon_{1,2}^{- *}
\propto \Gamma_{1,2} \epsilon_{1,2}^- \,,
\end{equation}
and at this stage we need to consider all possible combinations of
$1,2$ indices.
Since $\epsilon_1^-$ and $\epsilon_2^-$, and their complex
conjugations, always appear with $M_+$ and $M_-$ respectively,
flipping the index, say like $\epsilon_1^{- *} \leftrightarrow
\epsilon_2^-$, is not allowed.
So there remain two possibilities:

\paragraph{Case I:} $\Gamma_{1} \Bar\Gamma \epsilon_{1}^{- *} =\alpha_1
  \Gamma_{2} \epsilon_{1}^-  \,, \, \Gamma_{1} \Bar\Gamma
  \epsilon_{2}^{- *} =\alpha_2
  \Gamma_{2} \epsilon_{2}^- $ where $\alpha_{1,2} \in \bC$ \\
First by looking at the signature of $\epsilon_2^-$,
  $\alpha_1=\alpha_2$ is concluded, and thus we take $\alpha= \alpha_1
  = \alpha_2$ and obtain
\begin{align}
&  \hspace{-0.3cm}if (\Gamma - 1 )\epsilon = \notag \\
& \bigg[
-i \cos \eta T(\eta) 
-i 2 \pi \alpha' {\sin^2 \eta \over L^2} F_{\psi \rho} \alpha C(\eta) \sinh
 \rho
-i f T(\eta)
\bigg] 
\label{BPSeq1} \\[-1mm]
& \hskip2cm
\times 
(M_+ \epsilon_1^- + M_- \epsilon_2^-) \notag \\[2mm]&
+ \bigg[
-\cos \eta (\cosh \rho C(\eta) + \eta' \sinh \rho T(\eta))
+ f \cosh \rho C(\eta)
\bigg]
\label{BPSeq2} \\[-1mm]
& \hskip2cm
\times 
(M_+ \Gamma_{12} \epsilon_1^- - M_- \Gamma_{12} \epsilon_2^-) \notag\\[2mm]&
+
\bigg[
i \cos \eta ( \sinh \rho C(\eta) + \eta' \cosh \rho T(\eta))
+
i 2 \pi \alpha' { \sin^2 \eta \over L^2 } F_{\psi \rho }\alpha T(\eta)
-
i f \sinh \rho C(\eta)
\bigg]
\label{BPSeq3} \\& \hskip2cm
\times
(
M_+ g (\theta,\phi) \Gamma_1 \epsilon_1^-
+
M_- g (\theta,\phi) \Gamma_1 \epsilon_2^-
)
\notag \\[2mm]
& +
\bigg[
\eta' \cos \eta C(\eta)
+
2 \pi \alpha' { \sin^2 \eta \over L^2} F_{\psi \rho} \alpha C(\eta)
 \cosh \rho
\bigg] 
\label{BPSeq4} \\& \hskip2cm
\times
(
M_+ g(\theta,\phi) \Gamma_2 \epsilon_1^- 
-
M_- g(\theta,\phi) \Gamma_2 \epsilon_2^-
)
\notag \\
& =0 \,.
\end{align}
Each coefficient of $\epsilon$ terms has to vanish independently, and
by eliminating $F_{\psi\rho}$ and $f(\rho)$ from these four equations
we have
\begin{equation}
  \eta' \cot \eta = \coth \rho \,,
\end{equation}
which can be integrated to be
\begin{equation}
  \sin \eta = \kappa^{-1} \sinh\rho.
\end{equation}
Here $\kappa^{-1}$ is a constant of integration.
By inserting this solution for the condition 
which follows from the term (\ref{BPSeq4}), we have
\begin{equation}
  F_{\psi\rho}=
- \alpha^{-1} \frac{\kappa L^2}{2\pi \alpha' \sinh^2\rho}
\,,
\end{equation}
and also from (\ref{BPSeq2}), we have
\begin{equation}
  f(\rho)=\sqrt{1+\frac{1}{\kappa^2}(1+\alpha^{-2})}=1 \,,
\end{equation}
which leads $\alpha=\pm i$.
It is easy to see that these solve all conditions.

Thus a half BPS solutions in this case are
\begin{align}
  \sin \eta =& \kappa^{-1} \sinh\rho\,,
\\
F_{\psi\rho}=&
\pm i \frac{\kappa L^2}{2\pi \alpha' \sinh^2\rho}\,,
\\
\Gamma_1 \Bar\Gamma \epsilon_{1,2}^{* -} 
=& \pm i \Gamma_2 \epsilon_{1,2}^-  \,, 
\end{align}
where the signatures are taken to be same.
The solution found in \cite{Drukker:2005kx} corresponds to the plus sign.
In order to determine $\kappa$, one needs to solve the
equations of motion for the D3-brane action, and we have already known
that these BPS solutions solve the equations of motion as well.

\paragraph{Case II:} $\Bar\Gamma \epsilon_{1}^{- *} =\alpha_1
  \epsilon_{1}^- \,, \, \Bar\Gamma \epsilon_{2}^{- *} =\alpha_2
  \epsilon_{2}^-$ where $\alpha_{1,2} \in \bC$ \\
The analysis goes in parallel with the case I, and one finds
\begin{equation}
  f(\rho) = -\cos \eta \,,
\end{equation}
which does not have the solution since $0 \leq \eta \leq \pi/2$.
So this projection does not provide a BPS solution.

We therefore conclude that the BPS condition for D3-brane solutions
with the S$^2$ symmetric ansatz is sufficient to determine
the classical solution.
In the main part of the text, we have implicitly assumed that as for a
small deformation of the boundary condition $k \rightarrow k+p$ there
exists a unique classical solution associated with the new boundary
condition.
The result of this appendix justifies this prescription, since the
solution depends only on the parameter $\kappa$.


\end{document}